\documentclass[preprintnumbers,amsmath,amssymb]{revtex4}

\usepackage[dvips]{color}
\usepackage{url}
\definecolor{Red}{rgb}{1.00, 0.00, 0.00}

\usepackage{graphicx}
\usepackage{dcolumn}
\usepackage{bm}

\newcommand{\be}{\beta}

\newcommand{\de}{\delta}

\newcommand{\eps}{\epsilon}

\newcommand{\ka}{\kappa}
\newcommand{\la}{\lambda}

\newcommand{\si}{\sigma}

\renewcommand{\th}{\theta}   

\newcommand{\<}{\langle} 
\renewcommand{\>}{\rangle} 
\newcommand{\txt}{\textstyle}
\newcommand{\dsp}{\displaystyle}
\newcommand\eqn[1]{(\ref{#1})}      
\newcommand{\beq}{\begin{equation}}
\newcommand{\eeq}{\end{equation}}
\newcommand{\ba}{\begin{array}}
\newcommand{\ea}{\end{array}}
\newcommand{\bea}{\begin{eqnarray}}
\newcommand{\eea}{\end{eqnarray}}
\newcommand{\bi}{\begin{itemize}}  
\newcommand{\ei}{\end{itemize}}
\newcommand{\ben}{\begin{enumerate}} 
\newcommand{\een}{\end{enumerate}}

\newcommand{\half} {{\txt \frac{1}{2}}}

\newcommand{\vA}{{\bf A}}  
\newcommand{\vevp}{\langle \phi_p \rangle}
\newcommand{\vevn}{\langle \phi_n \rangle}
\newcommand{\rt}{\tilde{r}}

\newcommand{\subhalf} {{\frac{1}{2}}}
\newcommand{\feyn}[1]{
  \setbox0=\hbox{\ensuremath{#1}}
  \hbox to\wd0{\hbox to0pt{\hbox to\wd0{\hss/\hss}\hss}\box0}}

\begin{document}

\preprint{}

\title{
Flux tubes and the type-I/type-II transition in
a superconductor coupled to a superfluid
}

\author{Mark G. Alford and Gerald Good}
\affiliation{Physics Department, Washington University,
St.~Louis, MO~63130, USA}

\date{30 May 2008}

\begin{abstract}

We analyze magnetic flux tubes at zero temperature in a superconductor
that is coupled to a superfluid via both density and gradient
(``entrainment'') interactions. The example we have in mind is
high-density nuclear matter, which is a proton superconductor and a
neutron superfluid, but our treatment is general and simple, modeling
the interactions as a Ginzburg-Landau effective theory with
four-fermion couplings, including only $s$-wave pairing.
We numerically solve the field equations for
flux tubes with an arbitrary number of flux quanta, and compare their
energies.  This allows us to map the type-I/type-II transition in the
superconductor, which occurs at the conventional $\ka\equiv\la/\xi =
1/\sqrt{2}$ if the condensates are uncoupled.  We find that a density
coupling between the condensates raises the critical $\ka$ and,
for a sufficiently high neutron density, resolves the type-I/type-II
transition line into an infinite number of bands
corresponding to ``type-II(n)'' phases, in which
 $n$, the number of quanta in the
favored flux tube, steps from 1 to infinity. For lower
neutron density, the coupling creates spinodal regions around
the type-I/type-II boundary, in which 
metastable flux configurations are possible.
We find that a
gradient coupling between the condensates lowers the critical $\ka$
and creates spinodal regions.  
These exotic phenomena may 
not occur in nuclear matter, which is thought to be deep in the
type-II region, but might be observed in condensed matter systems.

\end{abstract}



\maketitle

\section{Introduction}

Superconductivity and superfluidity are well-studied phenomena, known to occur
in many physical systems, from cold metals and cold atomic gases to
nuclear matter and quark matter.
In this paper we investigate a system that has both a charged condensate,
leading to superconductivity, and a neutral condensate, leading to
superfluidity. We focus on the magnetic flux tubes that are associated
with the superconducting condensate, and study how they are modified by
the presence of the superfluid, assuming that the two condensates
can interact with each other via density and gradient (``entrainment'')
interactions.

An example of this type of system is nuclear matter, which at sufficiently 
high density undergoes Cooper pairing of both neutrons and protons.
We will present our calculations in this context, referring to the
charged condensate as the ``proton condensate'' and the neutral one
as the ``neutron condensate'', and choosing values  appropriate
to nuclear matter for our parameters when presenting numerical results.
In fact, the questions that we study in this paper were originally raised in 
investigations of the nature of the proton superconductivity in the 
nuclear matter in
a neutron star. Although it is generally believed that the protons
form a type-II superconductor \cite{BPP}, 
there is evidence from long neutron star
precession periods that seems to favor type-I superconductivity
\cite{blink} (for contrary views see \cite{Jones:2004xa,Sedrakian:2004yq}).  
This led Buckley~et.~al.~\cite{Zhitnitsky} to suggest that, if the
density interaction between the magnitudes of the neutron and proton Cooper
pair condensates is extremely strong, nuclear
matter would be a type I superconductor even if its penetration depth
$\la$ and coherence length $\xi$ obey the conventional condition
$\la/\xi > 1/\sqrt{2}$ for type II superconductivity. 
We have argued elsewhere that the assumption of a strong coupling
between the proton and neutron condensates is wrong for neutron star
matter \cite{Alford:2005ku}.  However, Buckley~et.~al.~were correct in
making the point that a superconductor will be affected by
interactions with a co-existing superfluid.

In this paper we study the type I versus type II nature of a 
(proton) superconductor
coupled to a (neutron) superfluid,
using an effective theory for the
protons and neutrons that contains four-fermion interaction terms
which lead to $s$-wave pairing. We do not include higher-angular-momentum
pairing, although
that would be needed for a more realistic analysis of high-density
nuclear matter.
Our analysis extends that
of Ref.~\cite{Zhitnitsky} in the following ways:
(a)~Our model, like that of Ref.~\cite{Zhitnitsky},
contains a coupling $a_{np}$ between the
magnitudes of the neutron and proton condensates, and
self-couplings $a_{nn}$ and $a_{pp}$, but 
we survey the whole range of values of $a_{np}$, from zero to
of order $a_{pp}$;
(b)~we also include ``entrainment''
interactions between the gradients of the proton
and neutron condensates;
(c)~we use a simpler and more direct method to 
study the type-I/type-II phase boundary, using
the energetics of flux tube coalescence/fission: we calculate
the energy of flux tubes with a wide range of magnetic fluxes,
from one quantum to several hundred quanta, and find which one
has the lowest energy per unit flux. As we will see, this has
the additional benefit of
allowing us to find exotic stable multi-quantum flux tubes, such as
have been found in systems of two coupled superconductors
\cite{Babaev:2004hk}. However, as we discuss below,
our analysis is not sensitive
to minima in the interaction energy at finite separation
between flux tubes.

Our analysis is entirely at zero temperature. This is 
a good approximation for neutron star matter near nuclear 
saturation density, where
the critical temperatures for the superfluid and superconductor
are of order MeV \cite{Dean:2002zx,Muther:2005cj,Fabrocini:2006xt}.
The temperature of a compact star drops below this value within minutes of 
its formation in a supernova, and is at or below 
the keV range after the first 1000 years \cite{Page:2005fq}.
When we discuss type-I versus type-II
behavior we are referring to the response of the system to a magnetic
field at the lower critical value, at $T=0$.

As far as we know, there has been no previous work on how
a flux tube in a superconductor is affected by
a gradient coupling to a co-existing superfluid.
However, there has been work on possible knot solitons \cite{Babaev:2002wa},
vortices in the $SO(5)$ model of high-temperature 
superconductivity \cite{MacKenzie:2003jp},
and on the complementary situation, 
a superfluid vortex with gradient coupling to a co-existing
superconductor.
There the coupling leads to the ``entrainment'' or Andreev-Bashkin effect
\cite{Andreev}
whereby the proton condensate is dragged along with the neutron
condensate, producing a non-zero proton current around
the vortex, dressing it with some magnetic flux \cite{Sedrakian:D1}. 
It is interesting to note
that this flux is not a multiple of the flux quantum for
proton flux tubes. 
This is possible because of the difference between the energetics of
a neutron vortex and a proton flux tube. The flux tube has energy
density localized to the vicinity of its core. Far from the
core the energy density must vanish, which means the proton
field must change in phase by a multiple of $2\pi$, and the
the vector potential must cancel the resultant gradient,
leading to a quantized magnetic flux. A neutron vortex,
by contrast, has gradient energy that is
not localized to the vicinity of the vortex, and the
total energy per unit length diverges
in the infinite volume limit. The vector potential
is therefore not constrained to cancel any gradient in the
proton field, and takes on a value that minimizes
the overall energy, with no quantization condition
on the resulting magnetic flux.

Returning to the situation that we study, a proton
flux tube in a neutron superfluid background, we do not expect
a similar behavior. This is because the proton flux tube's energy density
is localized around its core, giving it (unlike the neutron vortex)
a finite energy per unit length. If the neutron condensate were entrained,
and developed non-zero circulation around the flux tube, it would acquire
a non-localized energy density, leading to an infinite energy per unit
length for the flux tube, which is clearly energetically disfavored.
We will see below that the effect of gradient couplings on the 
proton superconductor is more subtle: it leads to metastable
regions near the type-I/type-II boundary.

\section{Stability of flux tubes}
\label{sec:stability}

Our aim is to explore the response of the proton superconductor 
to an applied critical magnetic field at zero temperature. We will
therefore construct a phase diagram in the space of
the coupling constants of the Ginzburg-Landau effective theory.
We would like to be able to specify when it is of type II
(at the lower critical magnetic field, flux tubes appear, and remain
separate, i.e they repel) and when it is of type I (at the
critical magnetic field, macroscopic normal regions appear,
i.e. the flux tubes attract and coalesce).
The simplest way to do this
is to calculate the energy per unit length $E_n$ of a flux tube 
containing $n$ flux quanta.
The same approach has been used for vortices in the
$SO(5)$ model \cite{Juneau:2001vz}.
It is convenient to 
work in terms of the energy per flux quantum,
\beq
  B_n = \frac{E_n}{n} - E_1 \ .
\eeq
When $B_n$ is negative the $n$-quantum 
flux tube is stable against fission into many single quantum flux tubes,
and it is energetically favorable for $n$ single quantum flux tubes
to coalesce into one $n$-quantum flux tube.
When $B_n$ is positive the $n$-quantum flux tube is unstable
against fission, and coalescence is energetically disfavored.
If one calculates $B_n$ for all $n$ then
the energetically favored value of $n$ is the one that
minimizes $B_n$.

In a traditional type I superconductor, 
small flux tubes attract each other and amalgamate into large 
ones and ultimately into macroscopic normal regions, so 
we would expect to find $B_n<0$ with its value dropping 
monotonically as $n$ rises. In a type II superconductor
we would expect $B_n>0$, with its value rising monotonically with $n$. 
Our calculations confirm
these results for a single superconductor, but
we will see that $B_n$ shows more complicated behavior
when the superconductor feels interaction with a co-existing
superfluid.

Calculations of $B_n$ are straightforward because they always
occur in a cylindrically symmetric geometry, so the problem
is one-dimensional. For a more detailed understanding of
flux tube interactions, one would have to
consider two single-quantum
flux tubes a distance $d$ apart. Their total energy is $U(d)$, where
$U(0)=E_2$ and $U(\infty)=2E_1$, so $B_2=\half(U(0)-U(\infty))$.
As expected, $B_2<0$ means that the flux tubes have lower energy
when they amalgamate, and  $B_2>0$ means that the flux tubes have lower energy
when they separate. If  $U(d)$ is monotonic, we can conclude that
flux tubes either coalesce ($B_2<0$) or repel to infinite separation
($B_2>0$), corresponding to type-I or type-II behavior respectively.
However, if there is a
minimum in $U(d)$ at some favored intermediate separation $d=d^*$
then irrespective of the sign of $B_n$, one has
a new variety of type II superconductor with some favored Abrikosov lattice 
spacing $d^*$. Such behavior has been found to arise from a $\phi^6$ term 
\cite{Mohammed} and in the case of
two charged condensates \cite{Babaev:2004hk}.
Calculating $U(d)$ in the current context
is an interesting but demanding problem
which we leave for future work. 
In this paper we assume that $U(d)$ is monotonic, so to
analyze the attractiveness/repulsiveness of the flux tube
interactions it is sufficient to calculate $B_n$, or equivalently
$E_n/n$.

\section{Flux tubes in the Ginzburg-Landau model}
\label{sec:eqns-of-motion}

\subsection{Ginzburg-Landau model}
\label{sec:GL}

We start by writing down the zero-temperature
Ginzburg-Landau effective theory of
proton and neutron condensates in the presence of a magnetic field
\cite{Alpar_et_al,Zhitnitsky}.
We denote the proton condensate field by $\phi_p$, the neutron condensate
field by $\phi_n$, and the magnetic vector potential by $\vA$. The free energy
density is
\beq
\label{eqn:FEdensity}
   {\cal F} = \frac{\hbar^2}{2 m_c} (|(\nabla - \frac{i q}{\hbar c}\vA)\phi_p|^2 + |\nabla \phi_n|^2)
  + \frac{|\nabla \times \vA|^2}{8\pi} + U_{ent}(\phi_p, \phi_n) + V(|\phi_p|^2,|\phi_n|^2)
\eeq
where $m_c$ is twice the nucleon mass, $q$ is twice the proton charge, 
$U_{ent}$ is the entrainment free energy density (see \cite{Alpar_et_al})
\bea
U_{ent} = -\frac{\hbar^2}{2m_c} \frac{\sigma}{2 \vevp \vevn} 
     \Bigl[\phi^*_p \phi^*_n \left((\nabla - \frac{i q}{\hbar c}\vA) \phi_p \cdot \nabla \phi_n \right) +
      \phi^*_p \phi_n \left((\nabla - \frac{i q}{\hbar c}\vA) \phi_p \cdot \nabla \phi^*_n \right)
      \notag \\
    + \phi_p \phi_n \left((\nabla + \frac{i q}{\hbar c}\vA) \phi^*_p \cdot \nabla \phi^*_n \right) 
    + \phi_p \phi^*_n \left((\nabla + \frac{i q}{\hbar c}\vA) \phi^*_p \cdot \nabla \phi_n \right) \Bigr]
\label{eqn:entrain}
\eea
and
\beq
  V(|\phi_p|^2,|\phi_n|^2)= -\mu­_p |\phi_p|^2 -\mu_n |\phi_n|^2 + \frac{a­_{pp}}{2} |\phi_p|^4 
  + \frac{a_{nn}}{2} |\phi_n|^4 + a_{pn} |\phi_p|^2 |\phi_n|^2
\label{eqn:potl}
\eeq
$\sigma$ is a parameter characterizing the strength of the gradient
coupling, $\mu_p$ and $\mu_n$ are the chemical potentials of the proton
and neutron condensate excitations, and $a_{pp}$, $a_{nn}$, and
$a_{pn}$ are the GL quartic couplings. 

In zero magnetic field, the condensates would have position-independent
bulk densities $\vevp^2$ and $\vevn^2$ obtained by minimizing the free energy.
This allows us to eliminate the chemical potentials
$\mu_p, \mu_n$ by writing
\bea
\mu_p &=& a_{pp} \vevp^2 + a_{pn} \vevn^2 \notag \\
\mu_n &=& a_{nn} \vevn^2 + a_{pn} \vevp^2
\eea
so up to constants involving $\vevp$ and $\vevn$, the potential $V$
can be expressed in terms of the deviations of the condensate fields
from their bulk values:
\beq
\ba{rl}
  V(|\phi_p|^2,|\phi_n|^2)= & \frac{a­_{pp}}{2} 
  \left(|\phi_p|^2-\vevp^2 \right)^2 
  + \frac{a_{nn}}{2} \left(|\phi_n|^2 -\vevn^2 \right)^2 \\ 
  &+ a_{pn} \left(|\phi_p|^2 -\vevp^2 \right) 
  \left( |\phi_n|^2 - \vevn^2 \right) \ .
\ea
\label{V-rel-to-bulk}
\eeq

\newcommand{\vv}{{\bf v}}
\newcommand{\vB}{{\bf B}}

In a neutron star, electrical neutrality keeps the proton fraction
small, in the 5\%\ to 10\%\ range \cite{Glendenning,APR98}; 
we will take $\vevp^2/\vevn^2\approx 0.05$.
As we now argue,
a typical value for the entrainment coupling is $\si\sim 10^{-1}$. 
We first relate our formalism to the hydrodynamic limit of the
free energy, following \cite{Alpar_et_al}. We focus on
the phases of the fields, $\phi_p = \vevp \exp(i\chi_p)$ and
$\phi_n =\vevn \exp(i\chi_n)$, and assume the fields have constant magnitude,
and their phases have gradients
\beq
\vv_p = \frac{\hbar}{2m_p} \nabla \chi_p - \frac{2e}{m_p c} \vA \ ,
\qquad
\vv_n =\dsp \frac{\hbar}{2m_n} \nabla \chi_n \ .
\eeq
The free energy density
 \eqn{eqn:FEdensity} then reduces to the hydrodynamic form
\beq
F = \frac{1}{2} \rho^{pp} \vv^2_p + \frac{1}{2} \rho^{nn} \vv^2_n 
  + \rho^{pn} \vv_p \cdot \vv_n + V + \frac{\vB^2}{8\pi} \ ,
\label{F-hydro}
\eeq
where the symmetric matrix $\rho$ of superfluid densities has elements
\beq
\rho^{pp} = 2m_p \vevp^2 \approx m_c \vevp^2 \ , \qquad
  ~\rho^{nn} = 2m_n \vevn^2 \approx m_c \vevn^2 \ , \qquad
  ~\rho^{pn} = -2m_n\sigma \vevp \vevn \ .
\label{rho}
\eeq
Our entrainment parameter $\si$ is therefore related to the parameter
$\eps$ of Ref.~\cite{Chamel:2006rc,Andersson:2002jd,Lindblom:1999wi} 
by $\si=\eps \vevn/\vevp$.
Since $\eps$ is of order 0.03, and $\vevn^2/\vevp^2 \sim 20$, 
we expect $\si\sim 10^{-1}$.
This is consistent with the estimate $\rho^{pn}\approx-\half \rho^{pp}$
used by \cite{Alpar_et_al}.
In terms of the Andreev-Bashkin parametrization \cite{Andreev},
$\rho_{12}=-\rho^{pn}$, $\rho_1 = \rho^{pp} + \rho^{pn}$,
$\rho_2 = \rho^{nn} + \rho^{pn}$,
so $\rho_1/\rho_{12} \sim 1$, $\rho_2/\rho_{12}\sim 40$.
All the interactions in \eqn{F-hydro}, including the entrainment, have
their ultimate origin in the strong interaction between the nucleons,
which is isospin symmetric, and hence does not distinguish protons
from neutrons.

\subsection{Flux tube solutions}

To study a flux tube containing $n$ flux quanta, 
we assume a cylindrically symmetric field configuration
in which the proton condensate field winds 
(in a covariantly constant way) around the $z$-axis with a 
net phase $2\pi n$,
\bea
\phi_p &=& \vevp ~f(r) e^{i n \th} \\
\phi_n &=& \vevn ~g(r) \\
\vA &=& \frac{n \hbar c}{q} \frac{a(r)}{r} \hat{\th} 
\eea
We have defined $\phi_n$ as a real field, because, as noted above,
any net phase change in the neutron condensate when it circles
the flux tube would cost an infinite energy per unit length.
Inserting the ansatz in (\ref{eqn:FEdensity}) we obtain 
\bea
{\cal F} &=& \frac{\hbar^2}{2 m_c} \left[ 
\vevp^2 \left( (f')^2 + \frac{n^2 f^2 (1-a)^2}{r^2} \right) + \vevn^2 (g')^2 
  - 2 \sigma \vevp \vevn f \cdot g \cdot f' \cdot g' \right] 
\notag \\ &~&~~~~~~~~+\frac{n^2\hbar^2 c^2}{8\pi q^2} \frac{(a')^2}{r^2}
+ \frac{a_{pp} \vevp^4}{2} \left( f^2 - 1 \right)^2
+ \frac{a_{nn} \vevn^4}{2} \left( g^2 - 1 \right)^2 \notag \\
&~&~~~~~~~~+ a_{pn} \vevp^2 \vevn^2 \left( f^2 - 1 \right)\left( g^2 - 1 \right)
\eea

Generating the Euler-Lagrange equations using the standard procedure, we obtain
a set of coupled differential equations for $f$, $g$ and $a$:

\bea
\frac{\hbar^2}{2 m_c a_{pp} \vevp^2} \Bigl[ f'' + \frac{f'}{r} - \frac{n^2 (1-a)^2 f}{r^2} 
    &-& \sigma \frac{\vevn}{\vevp} \bigl[ f\cdot g \left( g'' + \frac{g'}{r} \right) 
    + f \left( g' \right)^2 \bigr] \Bigr] \notag \\ 
&=& f(f^2-1) + \frac{a_{pn}}{a_{pp}} \frac{\vevn^2}{\vevp^2} f(g^2-1) \notag \\
\frac{\hbar^2}{2 m_c a_{pp} \vevp^2}\Bigl[ g'' + \frac{g'}{r} &-& \sigma \frac{\vevp}{\vevn} 
 \left[ f\cdot g \left( f'' + \frac{f'}{r}\right)+ g \left( f' \right)^2 \right] \Bigr] \notag \\
&=& \frac{a_{nn}}{a_{pp}} \frac{\vevn^2}{\vevp^2} g(g^2-1) + \frac{a_{pn}}{a_{pp}} g(f^2-1) \notag \\
\frac{m_c c^2}{4\pi q^2 \vevp^2} \left( a'' - \frac{a'}{r}\right) &=& - (1-a)f^2
\eea

At this point we recall the definition of the Ginzburg-Landau parameter $\kappa = \lambda / \xi$,
where the London penetration depth $\lambda$ and superconducting coherence 
length $\xi$ are (see \cite{Kittel})
\bea
\lambda &\equiv& \sqrt{\frac{m_c c^2}{4\pi q^2 \vevp^2}} = \sqrt{\frac{m_c c^2}{16\pi \hbar c \alpha_{EM} \vevp^2}} \notag \\
\xi &\equiv& \sqrt{\frac{\hbar^2}{2 m_c a_{pp} \vevp^2}}
\label{kappa}
\eea
To further simplify the equations, we then change variables to a dimensionless radial 
coordinate $\rt = r/\xi$, obtaining
\bea
\label{eqn:equations_of_motion}
f'' + \frac{f'}{\rt} - \frac{n^2 (1-a)^2 f}{\rt^2} 
    - \sigma \frac{\vevn}{\vevp} \left[ f\cdot g \left( g'' + \frac{g'}{\rt} \right) 
    + f \left( g' \right)^2 \right] 
&=& f(f^2-1) + \frac{a_{pn}}{a_{pp}} \frac{\vevn^2}{\vevp^2} f(g^2-1) \notag \\
g'' + \frac{g'}{\rt} - \sigma \frac{\vevp}{\vevn} 
 \left[ f\cdot g \left( f'' + \frac{f'}{\rt}\right)+ g \left( f' \right)^2 \right]
&=& \frac{a_{nn}}{a_{pp}} \frac{\vevn^2}{\vevp^2} g(g^2-1) + \frac{a_{pn}}{a_{pp}} g(f^2-1) \notag \\
a'' - \frac{a'}{\rt} &=& - \frac{1}{\kappa^2} (1-a)f^2
\eea

The free energy per unit length of the flux tube, in terms of the variable $\rt$, is
\beq
\label{eqn:free_energy}
\ba{rcl}
E_n &=&\dsp 2\pi a_{pp} \vevp^4 \xi^2 \int_0^\infty (\rt d\rt) \biggl\{  
  (f')^2 + \frac{n^2 f^2 (1-a)^2}{\rt} + \frac{\vevn^2}{\vevp^2} (g')^2 
  - 2\sigma \frac{\vevn}{\vevp} f \cdot g \cdot f' \cdot g' \\[2ex]
&&\dsp + n^2 \kappa^2 \frac{(a')^2}{\rt^2} 
   + \frac{1}{2} \left( f^2 - 1 \right)^2 + 
  \frac{1}{2} \frac{a_{nn}}{a_{pp}} \frac{\vevn^4}{\vevp^4} \left( g^2-1 \right)^2 +
  \frac{a_{pn}}{a_{pp}} \frac{\vevn^2}{\vevp^2} \left( f^2 - 1 \right) \left( g^2 - 1 \right)
   \biggr\} 
\ea
\eeq

In addition to the system of equations, we require boundary conditions
on the fields at the origin and at $\infty$. Far from the flux tube
core, the fields will go to their uniform condensate value, so
$f(\infty) = g(\infty) = a(\infty) = 1$. Near the origin,
 $f(r) \propto r^n$, $a(r) \propto r^2$ and $g(r)$ is a constant.
Therefore we have the conditions $f(0) = 0$, $a(0) = 0$ and $g'(0) =
0$.  To obtain the energy of a flux tube we numerically solve the ODE
system for the neutron and proton condensate and magnetic potential
profile functions, then calculate the free energy of the system by
inserting the results into \eqn{eqn:free_energy} and integrating.

The system has five independent parameters: $a_{pp}$, $a_{nn}/a_{pp}$,
$a_{pn}/a_{pp}$, $\sigma$, and $\vevn/\vevp$. In neutral nuclear matter,
the density of protons
(neutrons) is proportional to $\vevp^2$ ($\vevn^2$), and the proton
density is approximately 5\% of the total baryon number density
\cite{Alpar_et_al}, so we set $\vevp^2/\vevn^2 = .05$ in most of our
analysis.  
Following \cite{Zhitnitsky,Alford:2005ku} we set
$a_{nn} = a_{pp}$, and use \eqn{kappa} to exchange the
parameter $a_{pp}$ for $\kappa$, which is the conventional parameter
used in condensed matter studies of superconductivity.
Our reduced set of parameters is therefore $\kappa$,
the proton-neutron gradient coupling $\sigma$, and
the proton-neutron density coupling $\beta \equiv a_{pn}/a_{pp}$.
We also study some effects of varying $\vevp^2/\vevn^2$.

\begin{figure}[tb]
\begin{center} Profiles with nonzero density coupling \end{center}
 \includegraphics[width=0.48\textwidth]{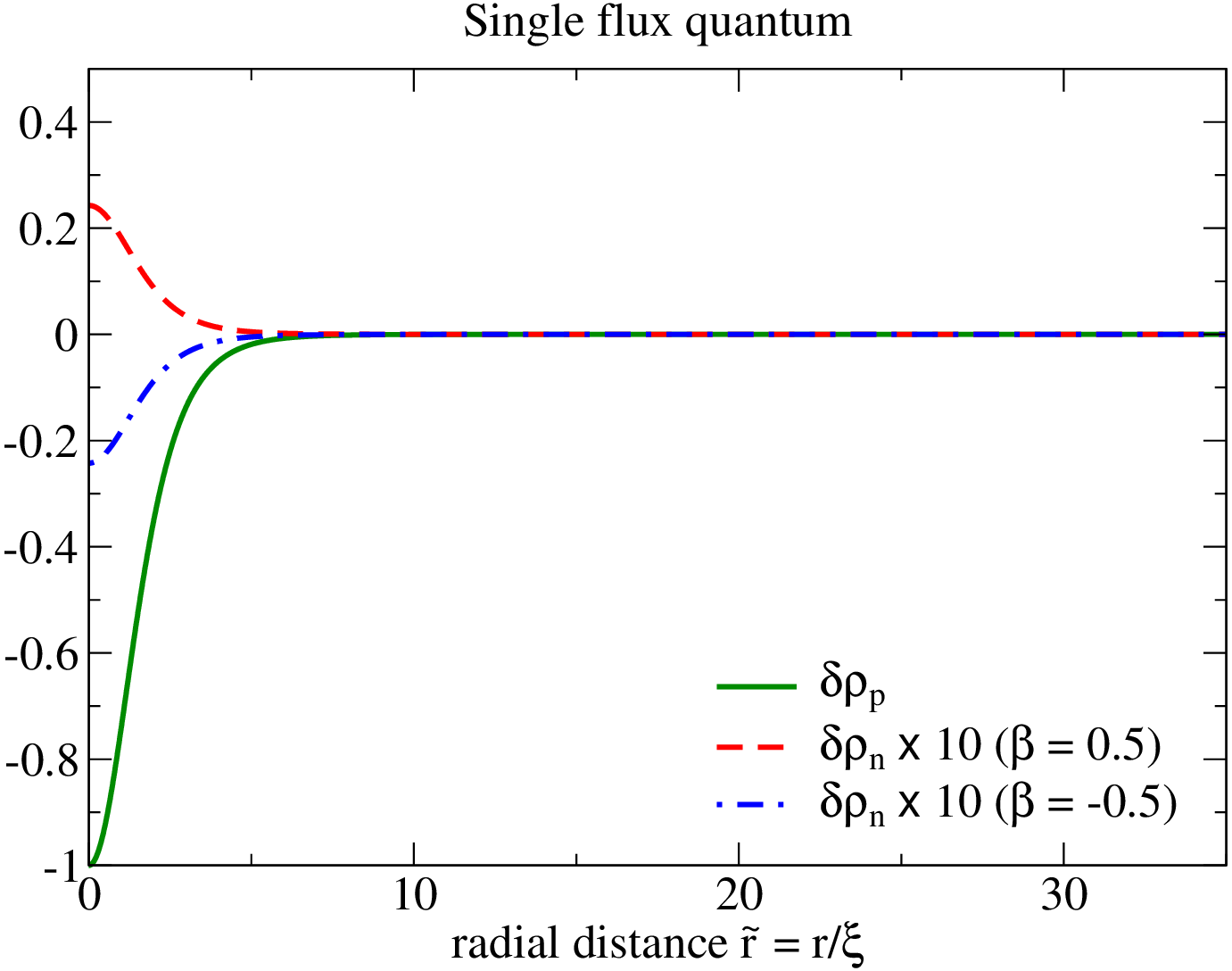}
 \includegraphics[width=0.48\textwidth]{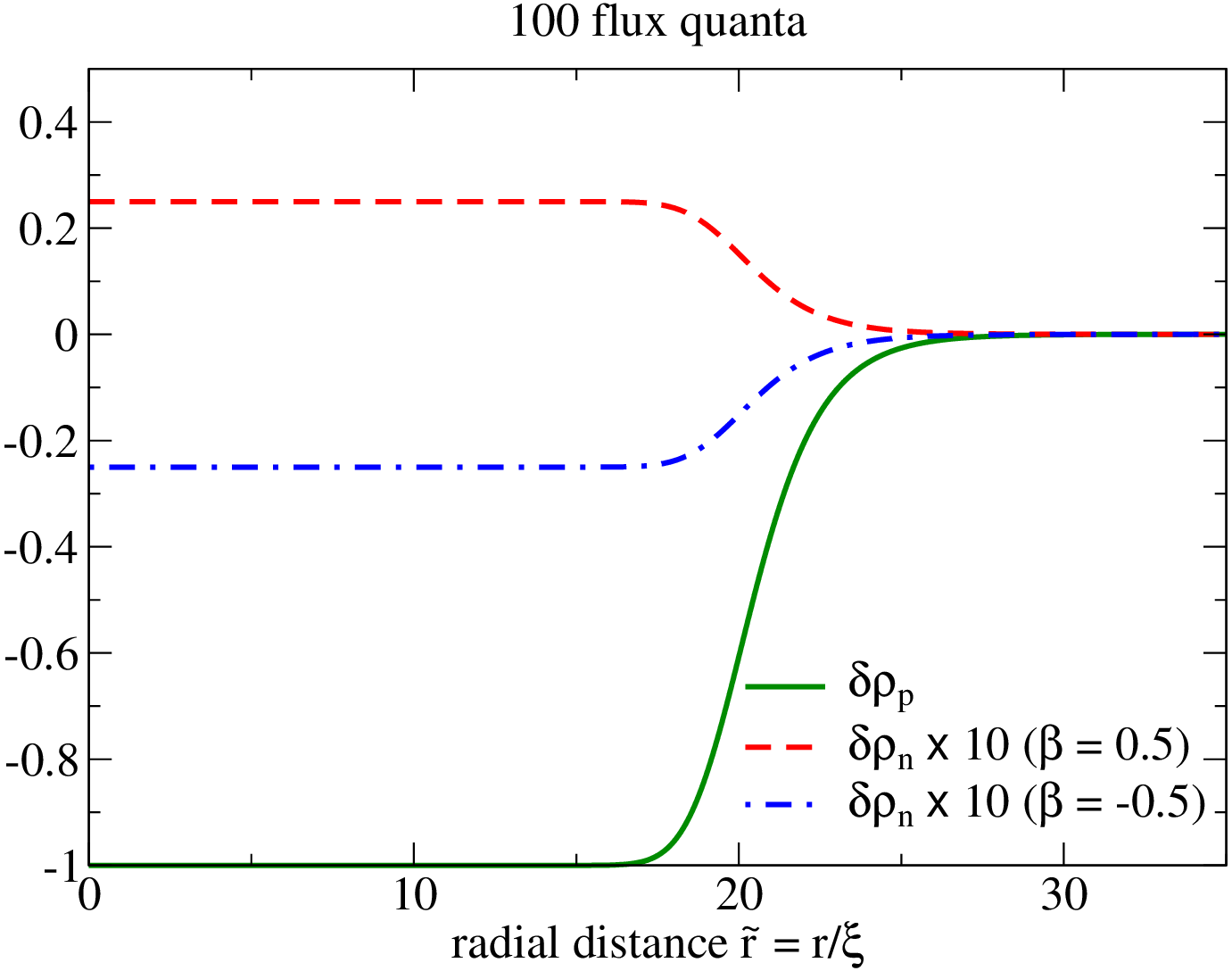}
\caption{
(Color online)
Profile of flux tube with $n=1$ units of flux (left) and $n=100$ units
of flux (right) showing the effect of density coupling $\beta$ between
neutron and proton condensates.  The plot shows the deviation
$\de\rho$ of the condensates from their vacuum values \eqn{delta_rho}.
With no coupling between the condensates ($\beta=\sigma=0$), the
neutrons are undisturbed ($\de\rho_n=0$). With a non-zero density
coupling $\be$, the neutron condensate (broken lines) is significantly
perturbed by the flux tube.  Note that the neutron $\delta \rho_n$'s
are multiplied by 10 (not by 100 as in Fig.~\ref{fig:profile-sigma}) to
make them visible.  The other parameters are $\kappa$ = 3.0, $\sigma$
= 0.0, and $\vevp^2/\vevn^2$=.05.
}
\label{fig:profile-beta}
\end{figure}

\begin{figure}[tb]
\begin{center} Profiles with nonzero gradient coupling \end{center}
 \includegraphics[width=0.48\textwidth]{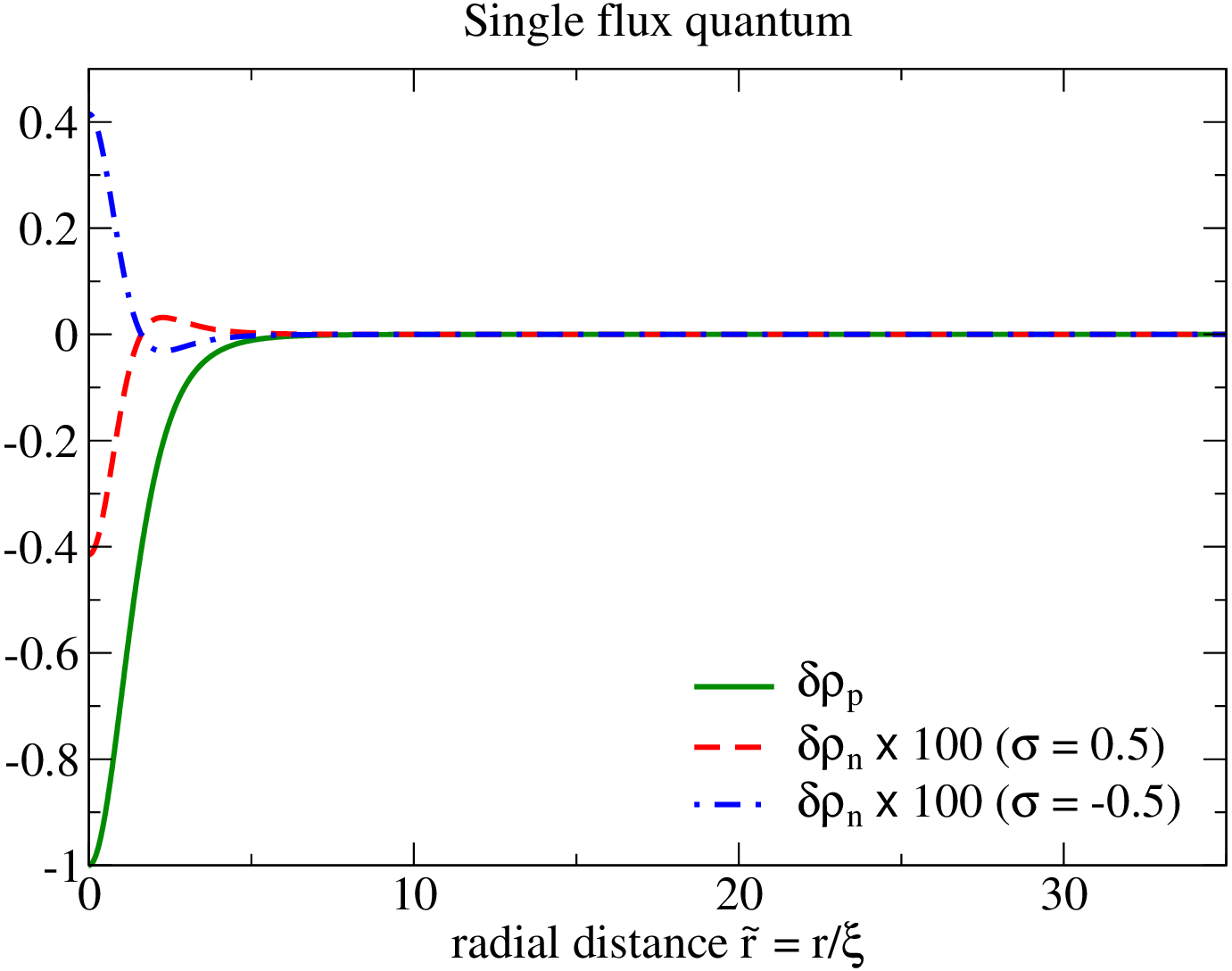}\hspace{0.04\textwidth}%
 \includegraphics[width=0.48\textwidth]{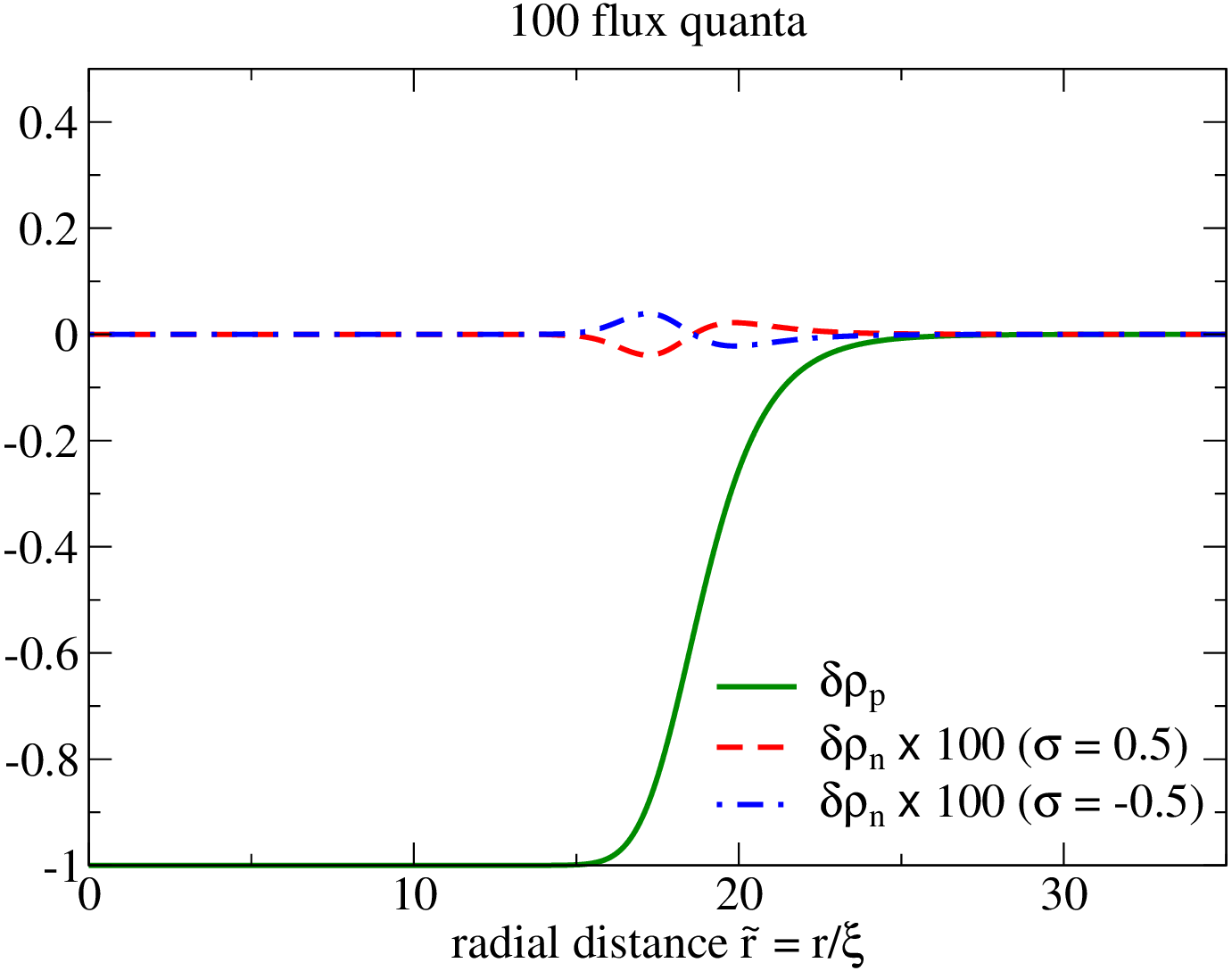}
\caption{
(Color online) 
Profile of flux tube with $n=1$ units of flux
(left) and $n=100$ units of flux (right) showing the effect of
gradient coupling $\sigma$ between neutrons and protons.
The plot shows the deviation
$\de\rho$ of the condensates from their vacuum values \eqn{delta_rho}.
With no coupling between the condensates ($\beta=\sigma=0$),
the neutrons are undisturbed ($\de\rho_n=0$). With a non-zero
gradient coupling $\sigma$,
the neutron condensate (broken lines) is 
slightly perturbed by the flux tube.
Note that the neutron $\delta \rho_n$'s are
multiplied by 100 to make them visible.
The other parameters are $\kappa$ = 3.0, $\beta$ = 0.0,
and $\vevp^2/\vevn^2$=.05.
}
\label{fig:profile-sigma}
\end{figure}

\section{Numerical Results}

\subsection{Flux tube solutions}

For given values of $\vevp^2/\vevn^2$, $\kappa$,
the proton-neutron gradient coupling $\sigma$,
and the proton-neutron amplitude coupling $\beta \equiv a_{pn}/a_{pp}$
we numerically solved the equations of motion
(\ref{eqn:equations_of_motion}) giving the field profiles for
flux tubes with various numbers $n$ of flux quanta.
We obtained the solutions using a finite-element relaxation method,
which is much less sensitive to initial conditions than the
traditional ``shooting'' method, and better suited to repeatedly
solving the equations for different sets of parameters.  Next, we
insert the solution for each profile into our expression for the
free energy (\ref{eqn:free_energy}) and numerically integrate it to
obtain a value for $E_n$.

To estimate the numerical errors in our results,
we varied the convergence criterion in the
finite-element relaxation calculation, the
spacing of the radial grid of points, and the radius out
to which the grid extended. 
We found that the resultant variation in $E_n/n$ was of order $10^{-6}$,
so numerical errors are invisible on the scale of the plots
shown in Fig.~\ref{fig:energyperflux}.

Having obtained $E_n$ we then plot the series $B_n$ to determine
whether the system is type I or type II for the chosen point in
parameter space. In this way we find the
points in parameter space where the system changes from a type I state
to a type II state. Taking various slices through the parameter space,
we can generate phase diagrams that show the boundary curves between
the various phases.

Figs.~\ref{fig:profile-beta} and \ref{fig:profile-sigma} each show a
profile for a flux tube with a single flux quantum $n=1$ on the left,
and a profile for a flux tube with 100 flux quanta on the right. 
Fig.~\ref{fig:profile-beta} shows the effect of non-zero density
coupling $\beta$ and Fig.~\ref{fig:profile-sigma} shows the effect of
non-zero gradient coupling $\sigma$. We
have plotted the normalized difference in density of the pair fields
from their condensate values,
\bea
\delta \rho_p (\rt) &\equiv& \frac{\phi^2_p (\rt) - \vevp^2}{\vevp^2} 
  = f^2 (\rt) - 1
\notag \\
\delta \rho_n (\rt) &\equiv& \frac{\phi^2_n (\rt) - \vevn^2}{\vevn^2} 
  = g^2 (\rt) - 1
\label{delta_rho}
\eea

\subsubsection{No coupling to neutrons}

We do not show a plot of the flux tube profile for a simple superconductor,
since this is well known: in a
core region whose area rises as the number of flux quanta $n$,
the proton condensate is suppressed;
in a wall region the condensate returns to its vacuum value.
At the Bogomolnyi point \cite{Bogomolnyi}, $\kappa = 1/\sqrt{2}$, 
the energy per flux
quantum is independent of $n$ \cite{deVega:1976mi},
but on either side of this value there
are area and perimeter contributions to the energy
\cite{Lukyanchuk}, so for $\ka$ close to $1/\sqrt{2}$ we expect the
energy of a flux tube in a simple superconductor to have the following
dependence on $n$, 
\beq
E^{(sc)}_n(\ka) =  n E_{\rm Bog}
  + \de\ka\, M \Bigl(n - c_\subhalf\sqrt{n} 
    + c_1 + \cdots \Bigr) \ .
\label{Esc}
\eeq
This is an expansion around $n=\infty$,
but our numerical results will show that it works down to $n=1$.
We define $\de\ka \equiv \ka-1/\sqrt{2}$. 
$E_{\rm Bog}$ is the energy per unit flux
at $\de\ka=0$. By convention we take the parameter
$M$, which has dimensions of energy, to be positive.
The value of $c_\subhalf$ is then positive, ensuring that for $\de\ka>0$,
$n=\infty$ is disfavored (type-II), and for $\de\ka<0$,
$n=\infty$ is  favored (type-I).
We will see this behavior in
our numerical results (Sec.~\ref{sec:En-none} and
upper left plot of Fig.~\ref{fig:energyperflux}).

\subsubsection{Density coupling to neutrons}
\label{sec:beta-coeffs}

For positive $\beta$, which corresponds to positive $a_{np}$,
equations (\ref{eqn:FEdensity})
and \eqn{eqn:potl} indicate that there is a repulsion between the neutron
and proton condensates, so in the center of the flux tube, where the
proton condensate is suppressed, the neutron condensate will be enhanced.
That is exactly what we see in Fig.~\ref{fig:profile-beta}, where the
dashed curve, showing the perturbation to the neutron density $\rho_n$,
rises inside the flux tube. For negative
$\be$ there is attraction between the two condensates, and the
neutron condensate is suppressed inside the flux tube (dash-dotted line).
We therefore expect that the leading correction due to the interaction
will be proportional to the core area, i.e proportional to $n$.
The energy of an $n$-quantum flux tube is then
\beq
E_n(\ka,\be) \approx E^{\rm (sc)}_n(\ka) 
 + M_\be(-n + b_\subhalf\sqrt{n} + b_1 + \cdots) \ ,
\label{eqn:dE-beta} 
\eeq
where $E^{(sc)}_n(\ka)$ is the energy for an $n$-quantum flux tube
in a pure superconductor, with no coupling to a superfluid
\eqn{Esc}.
The leading correction is $-M_\be n$, which should be
negative and quadratic in $\be$ for small $\be$
(see Sec.~\ref{sec:sign}), so
the interaction energy parameter $M_\be$ is positive and
proportional to $\be^2$.
The sub-leading term proportional
to $\sqrt{n}$ arises from the energy cost of
the gradient in $\rho_n$ at the edge of
the flux tube, where it must return to its vacuum value, so
we expect this term to be positive: $b_\subhalf>0$.
We do not have an {\em a priori} expectation for the sign of 
the sub-sub-leading term $b_1$.

\subsubsection{Gradient coupling to neutrons}

For positive $\si$, we expect from (\ref{eqn:FEdensity})
and \eqn{eqn:entrain} that the positive gradient in $\rho_p$
at the wall of the flux tube will induce a positive gradient
in $\rho_n$ in the same range of radii, which lowers the energy
of the system. This is exactly what
we see in Fig.~\ref{fig:profile-sigma}, where the dashed curve
showing the perturbation to $\rho_n$ has a positive slope in the
range of radii where the solid curve ($\rho_p$) has the largest
positive slope. On either side of that region it has a negative
slope, as it returns to its unperturbed value.
For negative $\si$ the effect is reversed:
the dash-dotted curve shows $\rho_n$ having a negative slope
where $\rho_p$ has the largest positive slope.

We therefore expect that in the presence of a gradient
coupling, the correction to the energy of a flux tube has a dominant
core-perimeter term proportional to $\sqrt{n}$, 
\beq
E_n(\ka,\si) \approx E^{\rm (sc)}_n(\ka) 
+ M_\si(-s_\subhalf\sqrt{n} + s_1 + \cdots) \ .
\label{eqn:dE-sigma} 
\eeq
The energy correction is negative and quadratic in $\si$ for small $\si$
(see Sec.~\ref{sec:sign}), so
the interaction energy parameter $M_\si$ is 
proportional to $\si^2$; choosing it to be positive by convention
requires $s_\subhalf$ to be positive.
We do not have an {\em a priori} expectation for the sign of $s_1$.

\subsubsection{Symmetry under change of sign of couplings}
\label{sec:sign}

It is clear from  Figs.~\ref{fig:profile-beta}
and \ref{fig:profile-sigma} that for couplings $\be$ and $\si$
of order $0.5$ the modification of the field configuration
due to the interaction between the condensates is extremely small, so
it is reasonable to treat its effects perturbatively.
(At the end of Sec.~\ref{sec:phase-diagrams} we will discuss the
limit of small neutron condensate, where the perturbative approach
becomes questionable.)

When we evaluate the perturbative correction to the
energy of the flux tube, there is no linear term
in $\be$ and $\si$. Such a term would arise from evaluating
the $\be$ and $\si$ terms from the Hamiltonian
in the unperturbed field configuration. But in that
configuration the neutron
condensate sits at its vacuum value, so both terms evaluate to zero
($g=1$, $g'=0$ in \eqn{eqn:free_energy}).

We therefore expect the change in the energy  of the flux tube
to be quadratic in the couplings $\be$ and $\si$.
Firstly, this correction must be negative. This is a well-known
result from perturbation theory: the second-order correction
arises from the change in the configuration in response to the perturbation,
which only occurs because it is driven by a resultant lowering of the energy.
Secondly, the change in the energy will in general contain
$\be^2$, $\si^2$, and $\be\si$ terms.
This means it will be even in $\be$ when $\si=0$ and
even in $\si$ when $\be=0$, so we expect
$M_\be \propto \be^2$ and $M_\si\propto\si^2$ in
Eqs.~\eqn{eqn:dE-beta} and \eqn{eqn:dE-sigma}.

However, if both 
$\be$ and $\si$ are nonzero, then the $\be\si$ terms spoil the symmetry
of the energy under negation of the couplings.
This is clear from Figs.~\ref{fig:profile-beta}
and \ref{fig:profile-sigma}.
For example, suppose that as well as non-zero $\be$
we have a very small non-zero $\si$.
Now consider sending $\be\to-\be$. From Fig.~\ref{fig:profile-beta}
we see that this changes the sign of the slope of $\rho_n$ in the 
wall region where $\rho_p$ has positive slope. If $\si$ is
nonzero then these two configurations will have different energies,
since the gradient of $\rho_n$ is then coupled to the gradient
of $\rho_p$.

\begin{figure}[tb]
\includegraphics[width=0.48\textwidth]{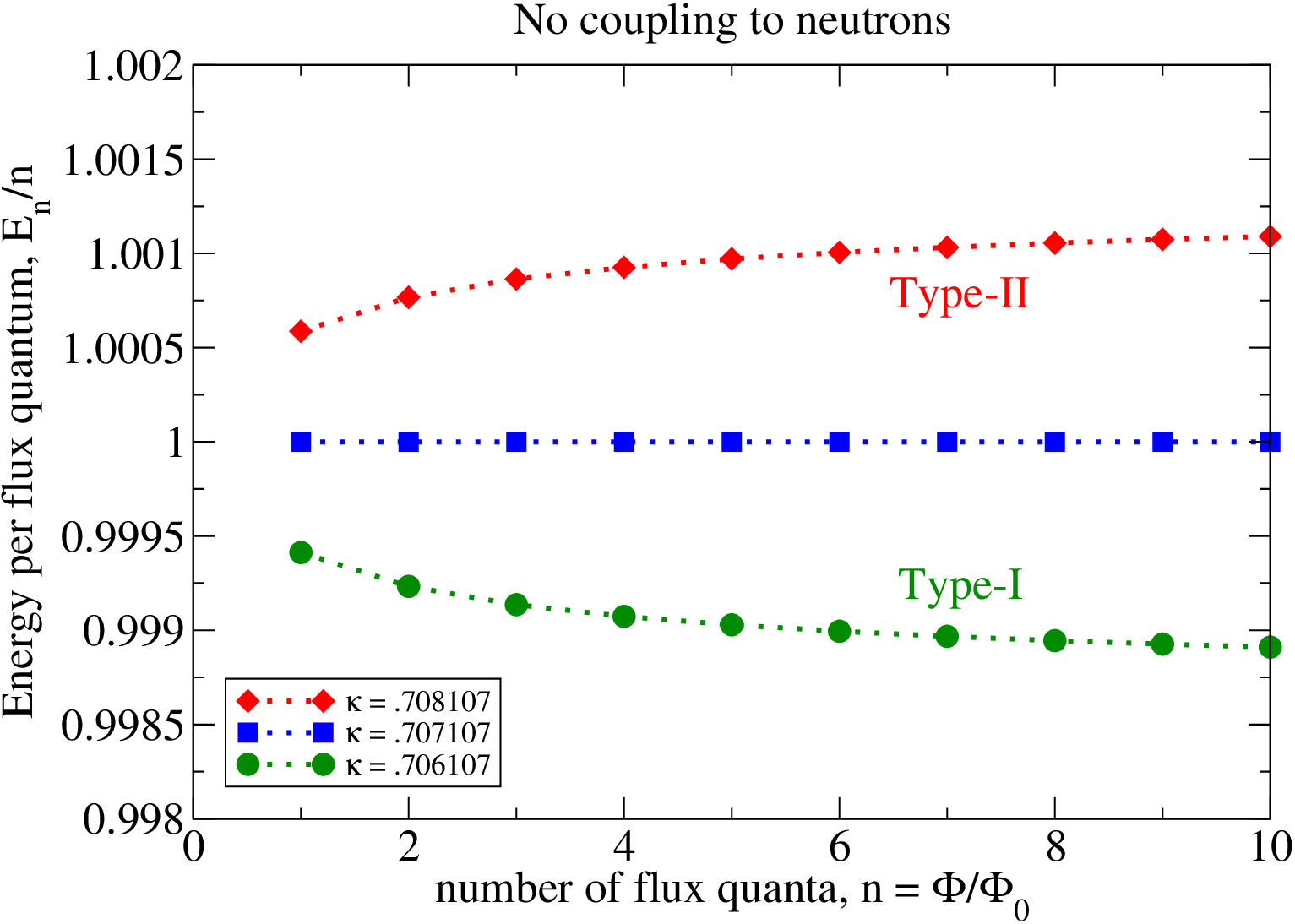}%
\hspace{0.04\textwidth}%
\includegraphics[width=0.48\textwidth]{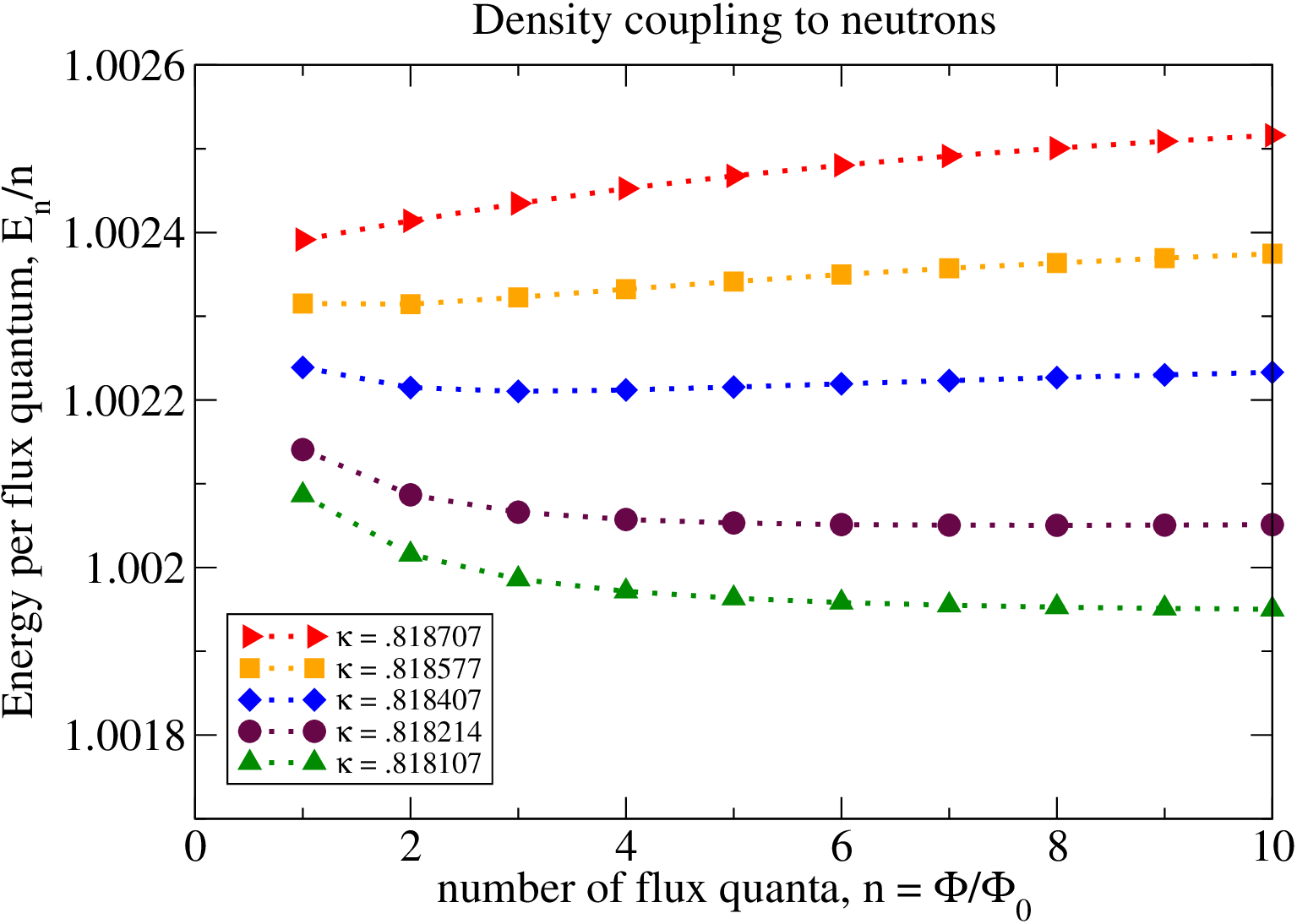}\\[2ex]
\includegraphics[width=0.48\textwidth]{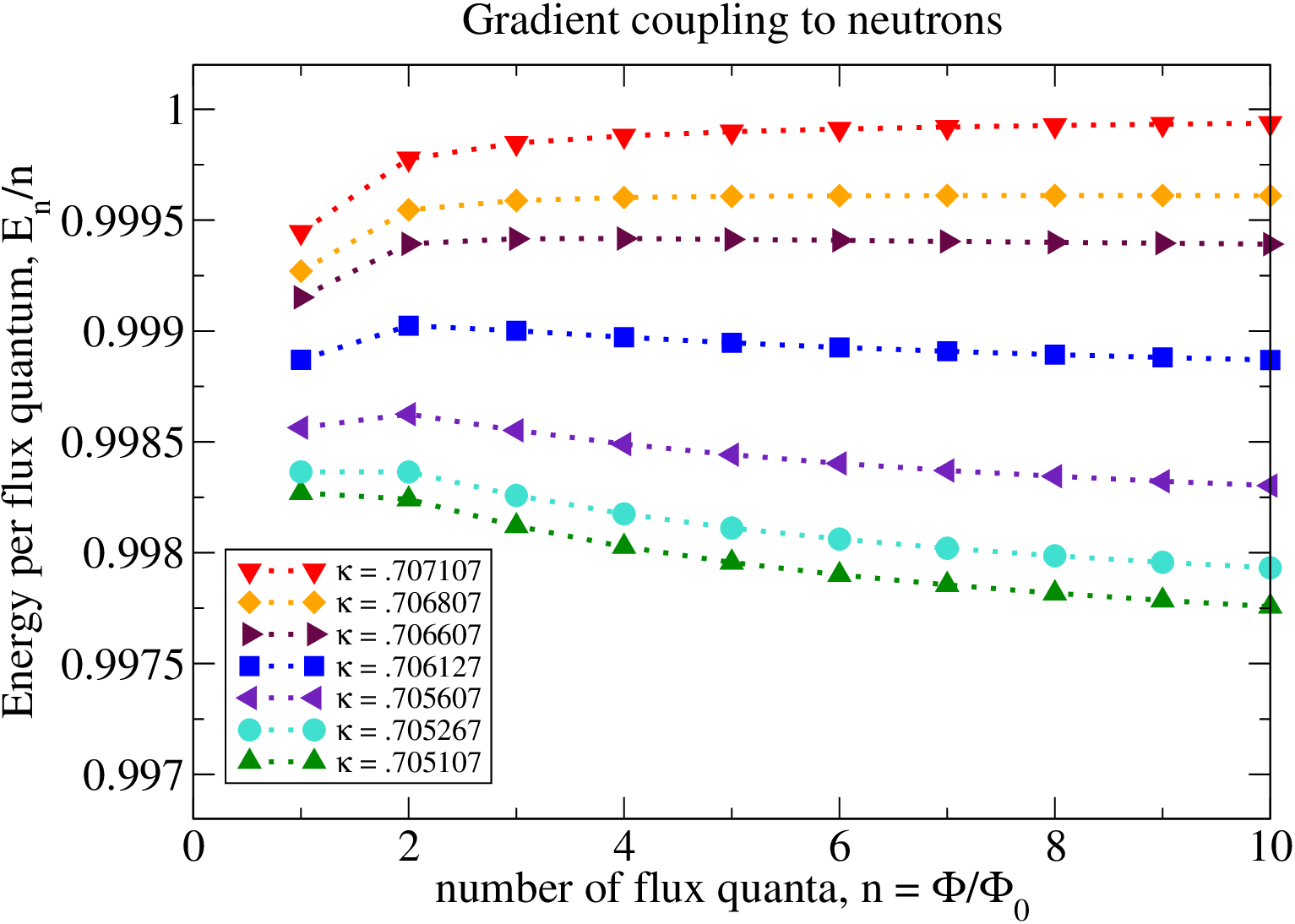}%
\hspace{0.04\textwidth}%
\includegraphics[width=0.48\textwidth]{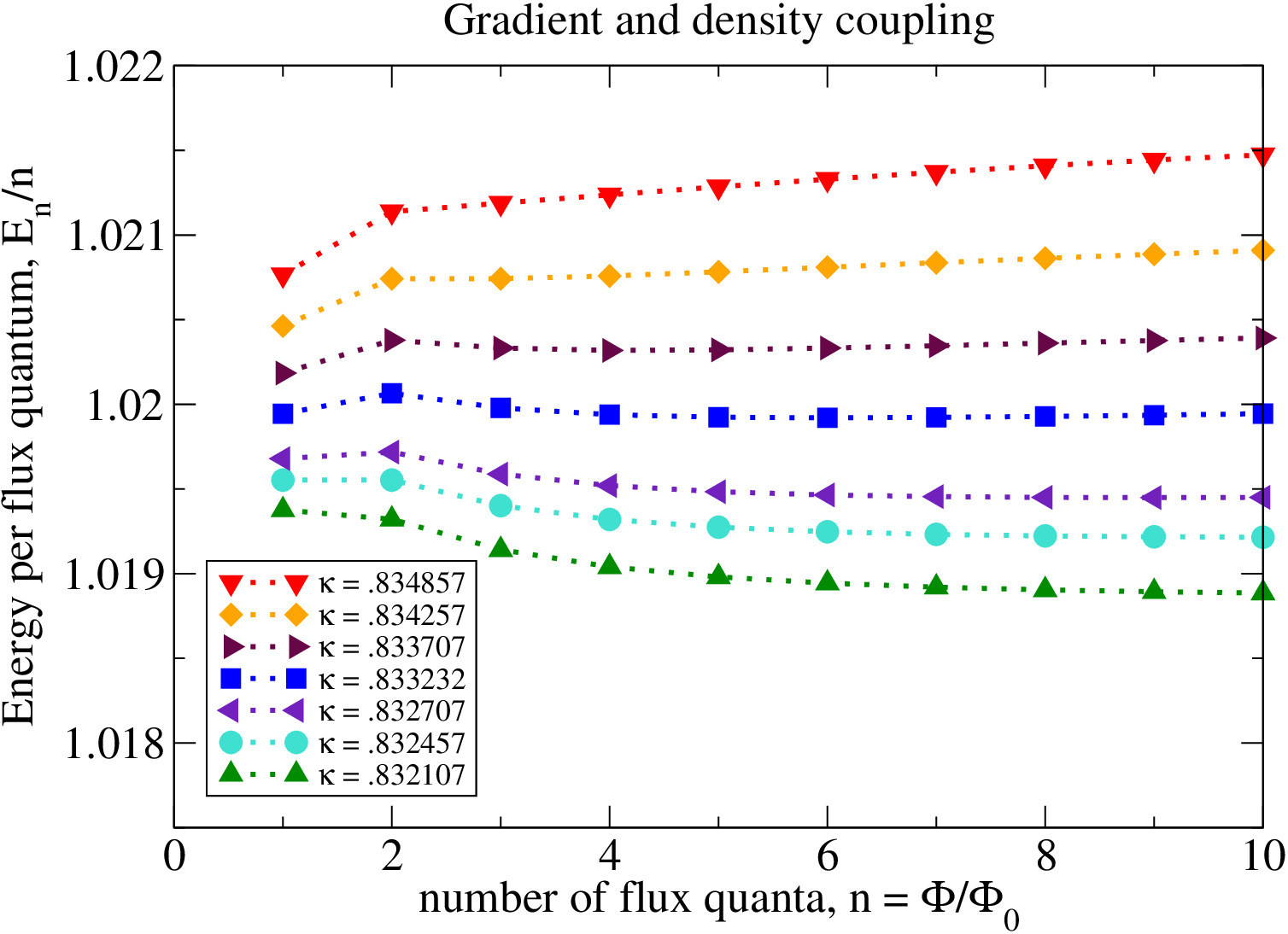}
\caption{
(Color online)
The energy per flux quantum $E_n/n$, in units of $E_{\rm Bog}$ 
(see Eq.~\eqn{Esc}),
as a function of the number $n$
of units of flux in the flux tube.
Top left, simple proton superconductor with neutrons completely
decoupled ($\beta = \sigma = 0$);
top right, density coupling between condensates ($\beta=.5,\sigma=0$);
bottom left, gradient coupling between condensates ($\beta = 0,\sigma=.5$); 
bottom right, both couplings ($\beta=\sigma=.5$).
}
\label{fig:energyperflux}
\end{figure}

\subsection{Energetic stability of flux tubes}

In Fig.~\ref{fig:energyperflux}, the energy per flux unit ($E_n/n$) is
plotted against $n$ for various values of the Ginzburg-Landau 
parameters, namely
$\kappa$, the density coupling $\beta$, and the gradient coupling $\sigma$.
We fixed $\vevp^2/\vevn^2=0.05$ (Sec.~\ref{sec:GL}).

\subsubsection{No coupling to neutrons}
\label{sec:En-none}

The upper left plot of Fig.~\ref{fig:energyperflux} shows 
$E^{(sc)}_n(\ka)/n$, the energy
per flux quantum when there are no
interactions between the neutron and proton pairs.
We see that the only possible
phases are the standard type I and type II, with a 
transition at the Bogomolnyi point, $\kappa =
1/\sqrt{2}$, where the favored value of $n$ jumps from 1 to infinity.. 
The lower line ($\kappa$ just below $1/\sqrt{2}$)
corresponds to type-I,
where the lowest energy/flux is at $n=\infty$, so flux tubes attract.
The upper line ($\kappa$ just above $1/\sqrt{2}$) corresponds to
type-II, where the lowest energy/flux is at $n=1$, so flux tubes 
always repel each other.
The middle line corresponds to the transition point ($\kappa=1/\sqrt{2}$),
where there is no interaction between flux tubes \cite{Bogomolnyi}.
Our numerical results are consistent with the expected
form  \eqn{Esc}:
when $\de\ka>0$ the asymptotic value
of $E_n/n$ is increased, and $E_n/n$ rises monotonically
towards that asymptotic value, and conversely
when  $\de\ka<0$ the asymptotic value
of $E_n/n$ is decreased, and $E_n/n$ falls monotonically
towards that asymptotic value. It is clear that $c_\subhalf$ 
in \eqn{Esc} must be positive to obtain this behavior at large $n$.
From fits to our numerical calculations we find that
$c_1$ is always positive, so it ``fights against'' the leading
$c_\subhalf/\sqrt{n}$ term, but for all $n\geqslant 1$ it is
overwhelmed.
In fact, we find that \eqn{Esc} gives an
excellent fit to our results down to $n=1$, without any higher order terms.

In the remaining panels of  Fig.~\ref{fig:energyperflux}, we
explore the effect of density and gradient couplings between the
proton superconductor and the neutron superfluid.

\subsubsection{Density coupling to neutrons}
\label{sec:En-density}

The upper right panel of Fig.~\ref{fig:energyperflux} shows the effect
of a density coupling between the condensates. 
From \eqn{Esc} and \eqn{eqn:dE-beta} we expect
\beq
E_n/n = E_{\rm Bog} + (M\de\ka - M_\be)
 + \frac{M_\be b_\subhalf - \de\ka M c_\subhalf}{\sqrt{n}}
 + \frac{M_\be b_1 + \de\ka M c_1}{n} + \cdots
\label{beta-En}
\eeq
The first point to notice is that the density coupling
shifts the critical $\ka$ to a larger value.
The transition between type-I and type-II occurs when the
asymptotic behavior at large $n$ changes from rising to falling,
i.e.~when the coefficient of the $1/\sqrt{n}$ term changes sign.
This occurs for some positive value of $\de\ka$
\beq
\de\ka_{\rm crit}(\be) = \frac{M_\be \, b_{\subhalf}}{M c_\subhalf}
\quad\propto \be^2
\label{dk-beta}
\eeq
which rises as $\be^2$ because
$M$, $M_\be$, $b_\subhalf$, and $c_\subhalf$ are all positive,
and $M_\be\propto\be^2$ when $\si=0$ (Sec.~\ref{sec:sign}).
Thus in the upper right panel of Fig.~\ref{fig:energyperflux}
we had to increase $\ka$ from around $0.707$ to
around $0.818$ in order to find the transition.

The other important point is the presence
of a minimum in $E_n/n$  when $\ka$ is just above the new
type-I/type-II boundary, indicating that the favored value
of $n$ may be neither 1 (standard type-II) nor infinity
(type-I) but some intermediate value.
This is consistent with
\eqn{beta-En}, as long as we assume that the coefficient $b_1$
from \eqn{eqn:dE-beta} is either positive,
or negative and of sufficiently small magnitude, 
so that the $1/n$ term in \eqn{beta-En} has a positive coefficient
(recall that $M_\be$, $M$, and $c_1$ are all positive,
and $\de\ka$ is also positive in this region).
The minimum will then arise from competition between the 
positive $1/n$ term, which dominates at smaller $n$, giving
a negative slope, and the $1/\sqrt{n}$ term which 
has a negative coefficient (because $\de\ka$ is just above the
new critical value) and dominates
at larger $n$ giving a positive slope. However, 
as $\de\ka$ is reduced the negative coefficient of $1/\sqrt{n}$
becomes smaller and smaller, and the minimum 
moves out to arbitrarily large $n$, so the energetically favored 
value of $n$ does not jump
suddenly from 1 to $\infty$ as in the standard case,
but increases in steps from 1 to
infinity as we lower $\ka$ through a range of values
down to the new critical value.  
This creates an infinite number of
``type-II(n)'' phases, each with a different flux in the favored flux tube,
and when that flux becomes infinite the superconductor becomes type-I.
This behavior is seen in our numerical results 
(Fig.~\ref{fig:phase-kappa-beta}).

\subsubsection{Gradient coupling to neutrons}
\label{sec:En-gradient}

The lower left panel of Fig.~\ref{fig:energyperflux} shows the effect
of a gradient interaction with the superfluid.
From  \eqn{Esc} and \eqn{eqn:dE-beta} we expect
\beq
E_n/n = E_{\rm Bog} + M\de\ka
 + \frac{-M_\si s_\subhalf - \de\ka M c_\subhalf}{\sqrt{n}}
 + \frac{-M_\si s_1 + \de\ka M c_1}{n} + \cdots
\label{sigma-En}
\eeq
Here we see that the gradient coupling
shifts the critical $\ka$ to a smaller value.
The transition between type-I and type-II occurs when
the coefficient of the $1/\sqrt{n}$ term changes sign, which in this
case happens for small negative $\de\ka$, 
\beq
\de\ka_{\rm crit}(\si) = -\frac{M_\si \, s_{\subhalf}}{M c_\subhalf}
\quad\propto -\si^2
\label{dk-sigma}
\eeq
which is proportional to $-\si^2$ because
$M$, $M_\si$, $s_\subhalf$, and $c_\subhalf$ are all positive,
and $M_\si\propto\si^2$ when $\be=0$ (Sec.~\ref{sec:sign}).

The other important feature of this plot is the presence
of a maximum in $E_n/n$  when $\ka$ is close to the
type-I/type-II boundary. This is consistent with
\eqn{sigma-En}, as long as we assume that the coefficient $s_1$
from \eqn{eqn:dE-sigma} is either positive,
or negative and of sufficiently small magnitude, 
so that the $1/n$ term in \eqn{sigma-En} has a negative coefficient.
The maximum will then arise from competition between the 
negative $1/n$ term, which dominates at smaller $n$, giving
a positive slope, and the $1/\sqrt{n}$ term, which dominates
at larger $n$ giving a negative slope.

The presence of this maximum allows for the possibility of
metastable flux configurations. If we scan down in $\ka$, we start in
a type-II region where $E_n/n$ has its minimum at $n=1$ and rises
monotonically with $n$.  But at some point a metastable minimum at
$n=\infty$ appears, which drops to become degenerate with the minimum
at $n=1$. At this point there is a first-order transition: at the
critical field, $n=1$ flux tubes would co-exist with macroscopic
normal regions (i.e.~flux tubes with $n=\infty$) but not with
flux tubes of intermediate size.  Reducing $\ka$
further, the $n=1$ flux tube becomes energetically metastable, and
finally unstable.

\subsubsection{Density and gradient coupling to neutrons}
\label{sec:En-both}

The lower right panel of Fig.~\ref{fig:energyperflux} shows the effect
of a combination of gradient and density interactions.  As $\ka$ is
decreased, a metastable energy minimum emerges at finite $n$; it drops
and becomes a new global minimum at $n=n^*$, yielding a sharp
transition from $n=1$ type-II to $n=n^*$ type-II. As $\ka$ is reduced
further the favored number of flux quanta in a flux tube rises in
integer steps from $n^*$ to infinity, at which point the
superconductor becomes type-I.

\begin{figure}[htb]
\includegraphics[width=0.47\textwidth]{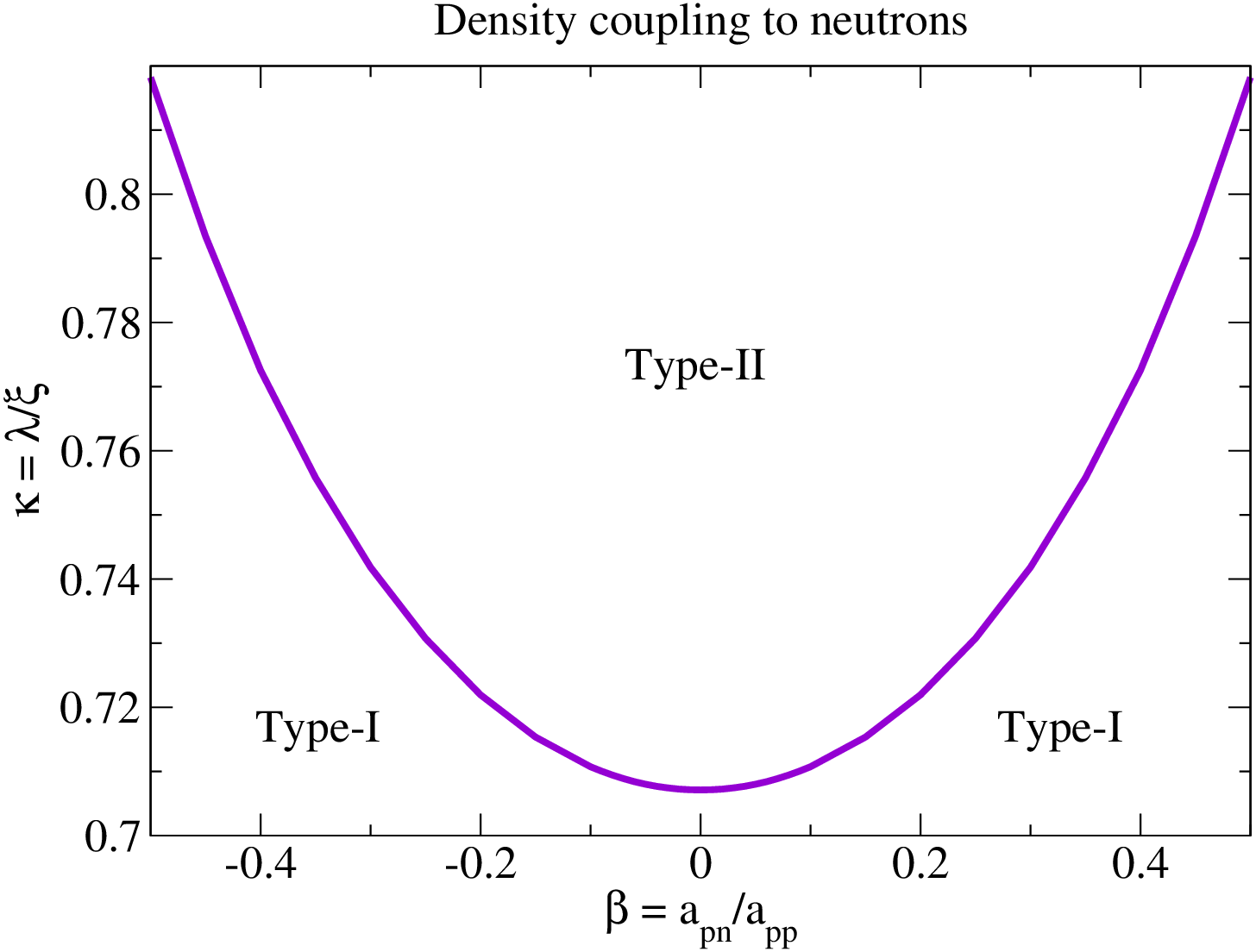}
\includegraphics[width=0.51\textwidth]{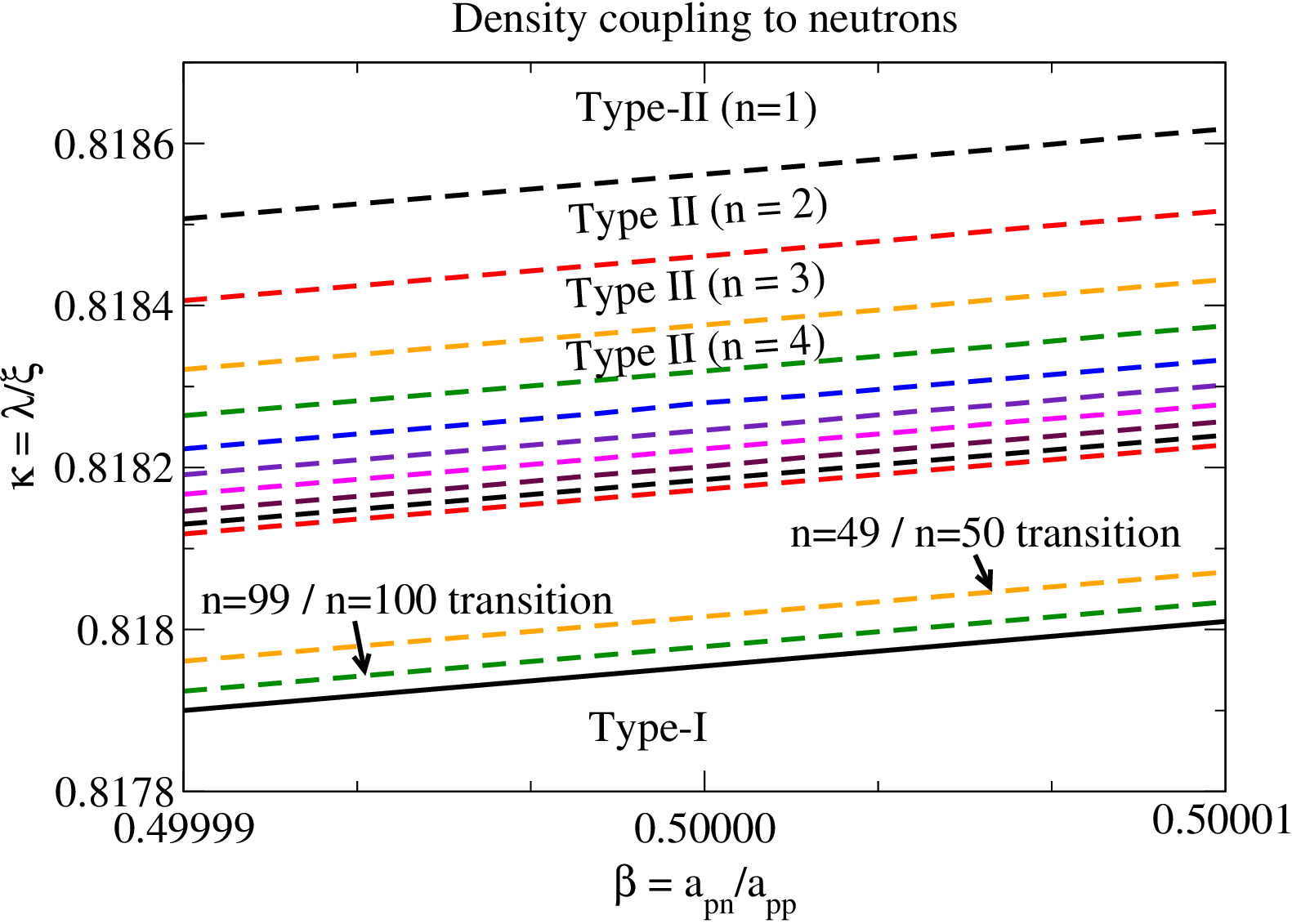}
\caption{
(Color online)
Effect on the superconductor
of density coupling $\beta$ to a superfluid, displayed as a
phase diagram in the $\kappa$-$\beta$ plane, with no gradient
coupling ($\si=0$) and $\vevp^2/\vevn^2=0.05$.
The left panel shows how non-zero $\be$ 
causes an increase in $\ka_{\rm critical}$.
In the right panel we magnify
the transition region near $\beta = 0.5$, illustrating that on the
type-II side there is a sequence of ``type-II(n)'' bands in which
the number of flux quanta in the favored flux tube
rises, reaching infinity when the superconductor becomes type I.
}
\label{fig:phase-kappa-beta}
\end{figure}

\begin{figure}[htb]
\includegraphics[width=0.48\textwidth]{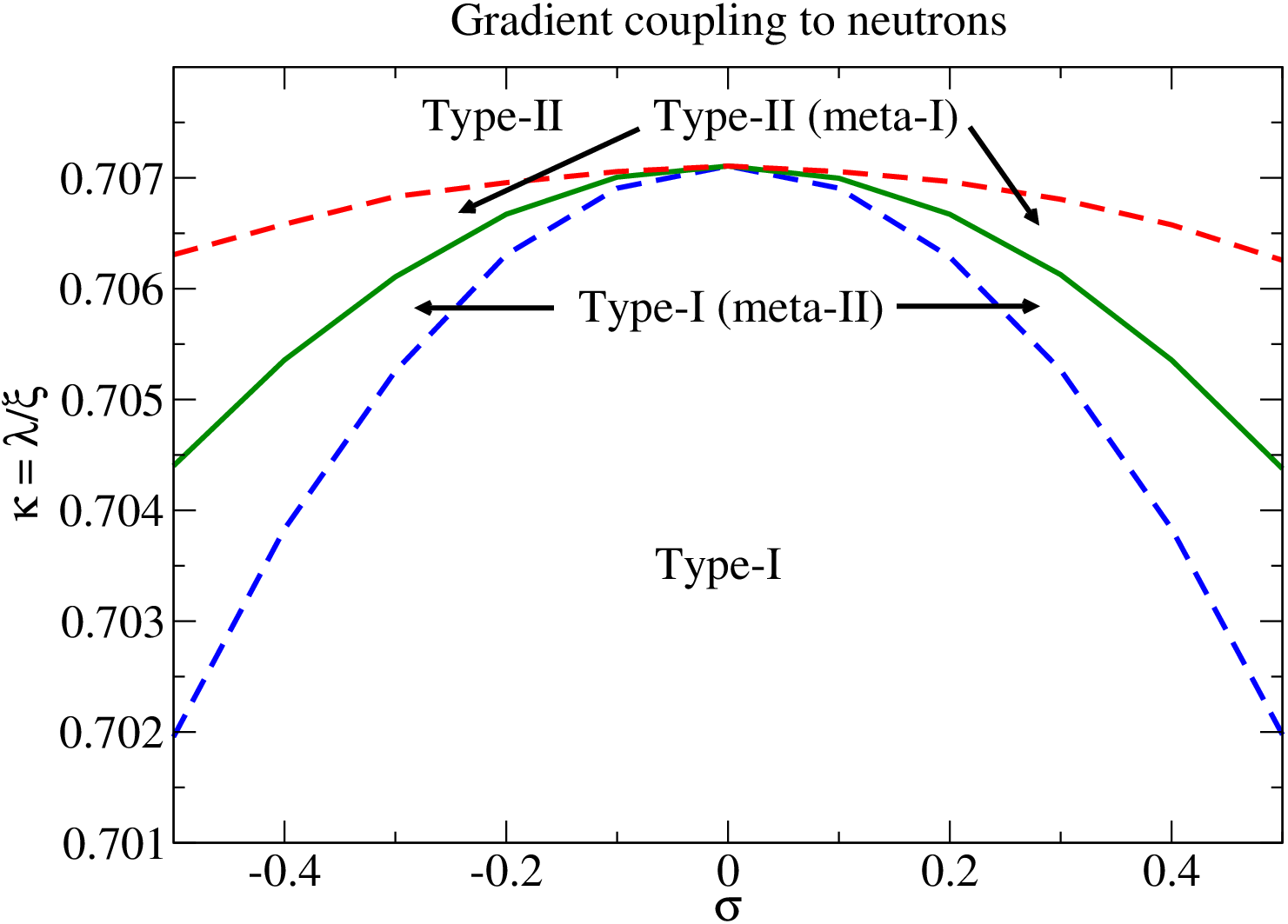}
\caption{(Color online)
Effect on the superconductor
of gradient coupling $\si$ to a superfluid, displayed as a
phase diagram in the $\kappa$-$\sigma$ plane, with no density
coupling ($\be=0$) and $\vevp^2/\vevn^2=0.05$. The gradient coupling  
causes a decrease in $\ka_{\rm critical}$, and
creates metastable states on either side of the transition,
with spinodal lines as shown.
\label{fig:phase-kappa-sigma}
}
\end{figure}

\begin{figure}[htb]
\includegraphics[width=0.46\textwidth]{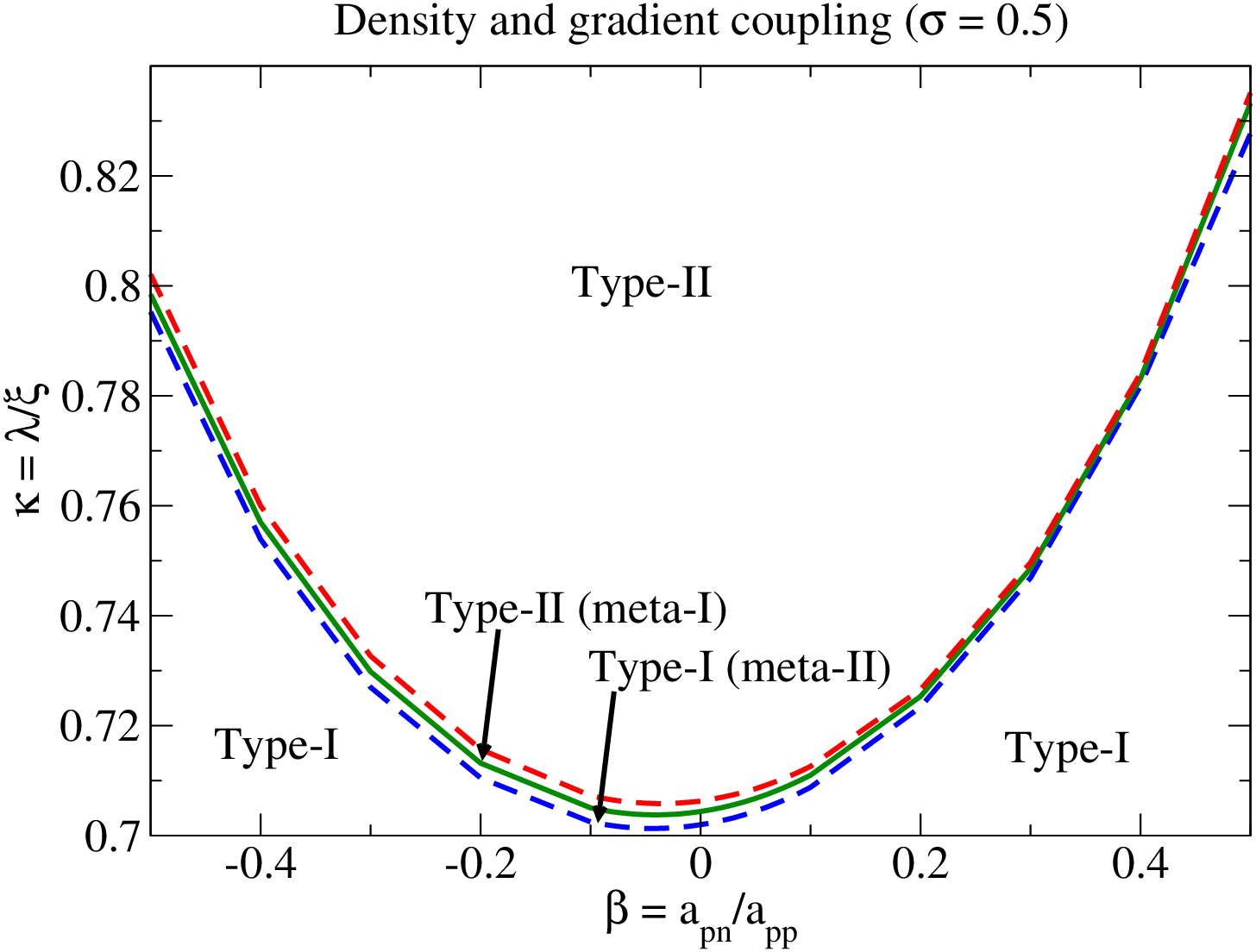}
\includegraphics[width=0.52\textwidth]{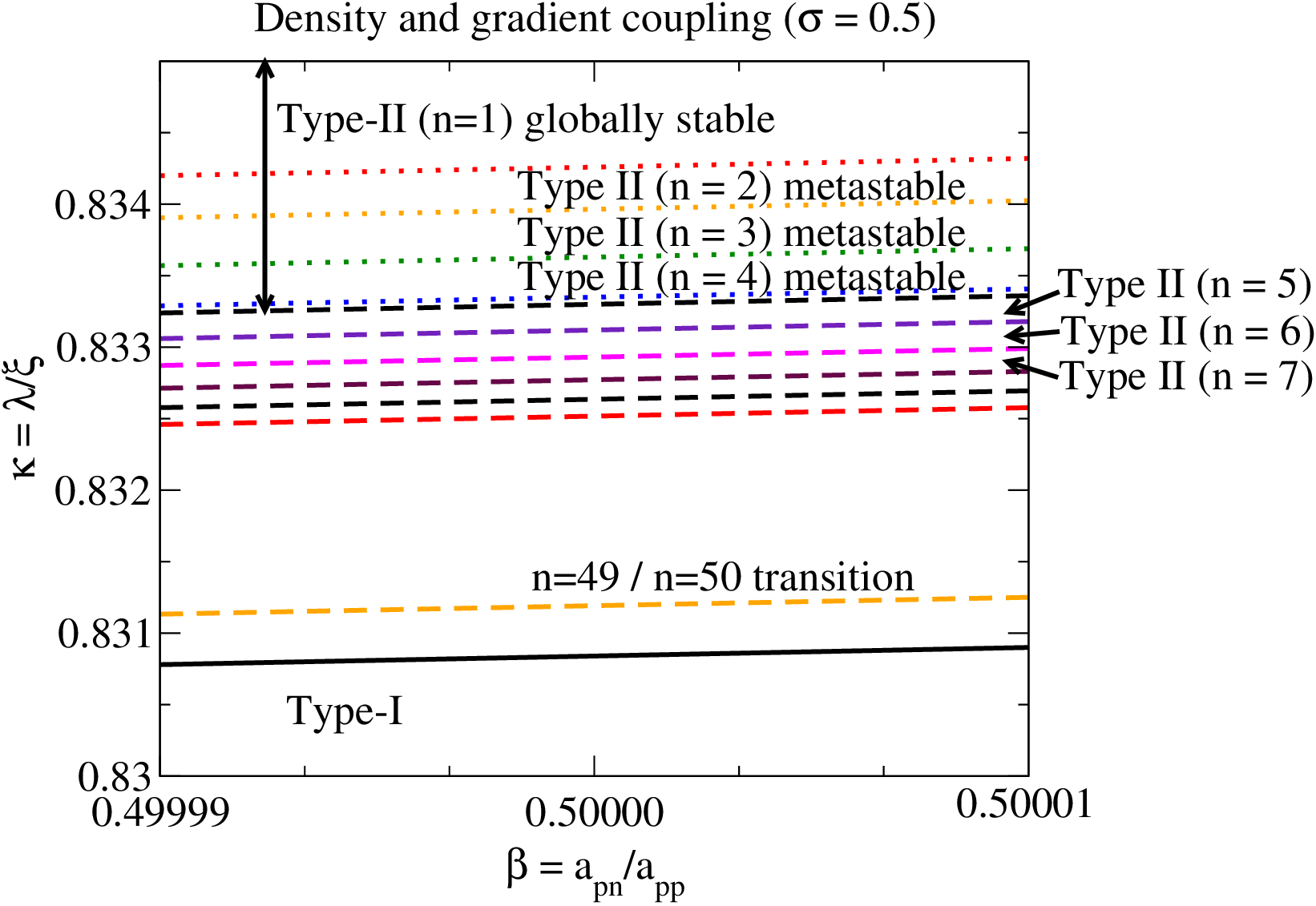}
\caption{(Color online)
Phase diagram for combined density and gradient interactions:
the $\kappa$-$\beta$ plane for $\sigma = 0.5$ and $\vevp^2/\vevn^2=0.05$. 
The type-I/type-II
boundary is no longer symmetric under $\be\to-\be$. 
In the right panel we magnify
the transition region near $\beta = 0.5$, illustrating that on the
type-II side as $\ka$ decreases 
the number of flux quanta in the favored flux tube jumps from
1 to a finite value (in this case $n=5$) 
and then there is a sequence of bands in which
$n$ rises to infinity, at which point the superconductor becomes type I.
}
\label{fig:phase-kappa-beta-sigma0.5}
\end{figure}



\begin{figure}[htb]
\includegraphics[width=0.48\textwidth]{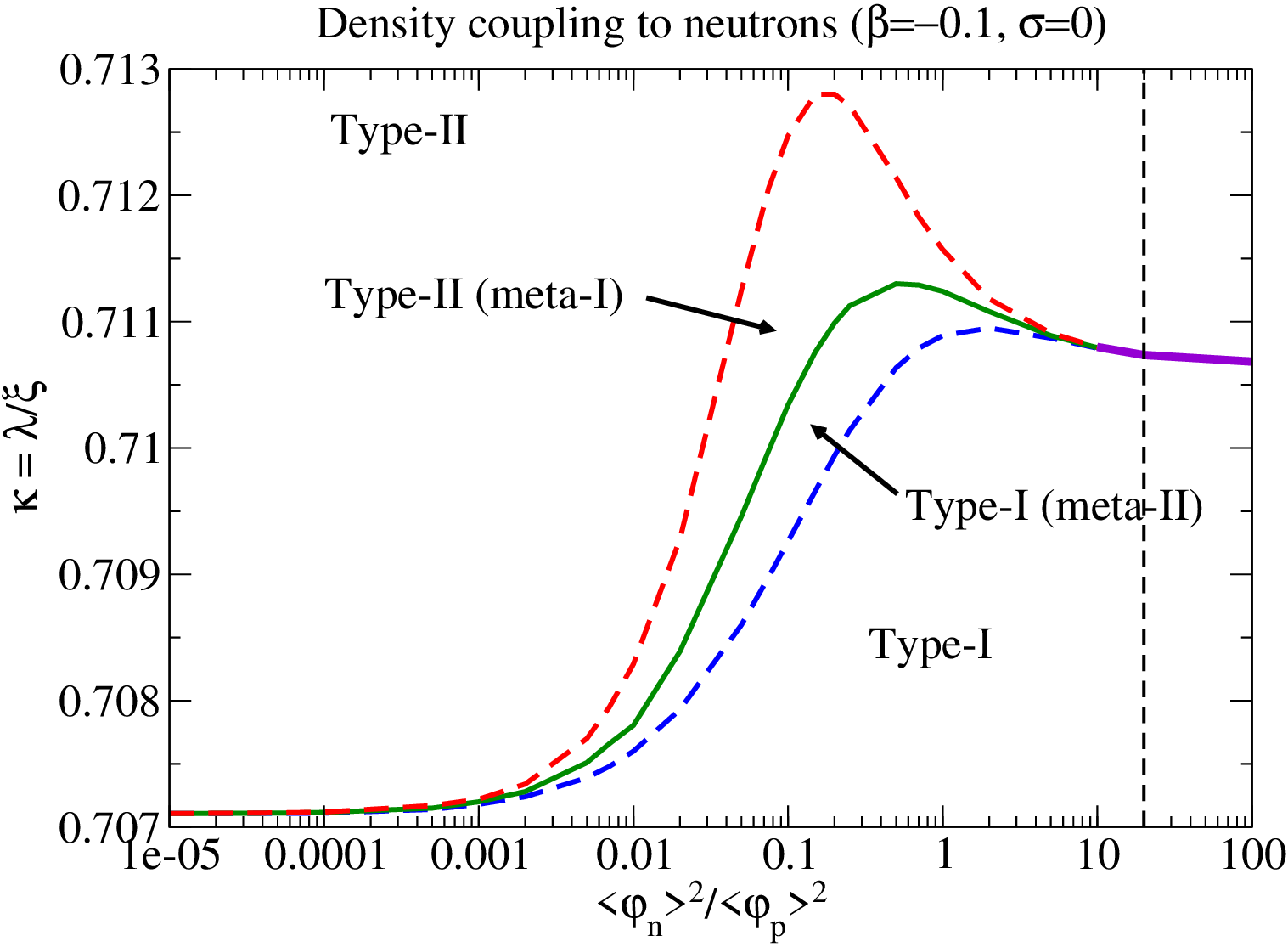}
\includegraphics[width=0.48\textwidth]{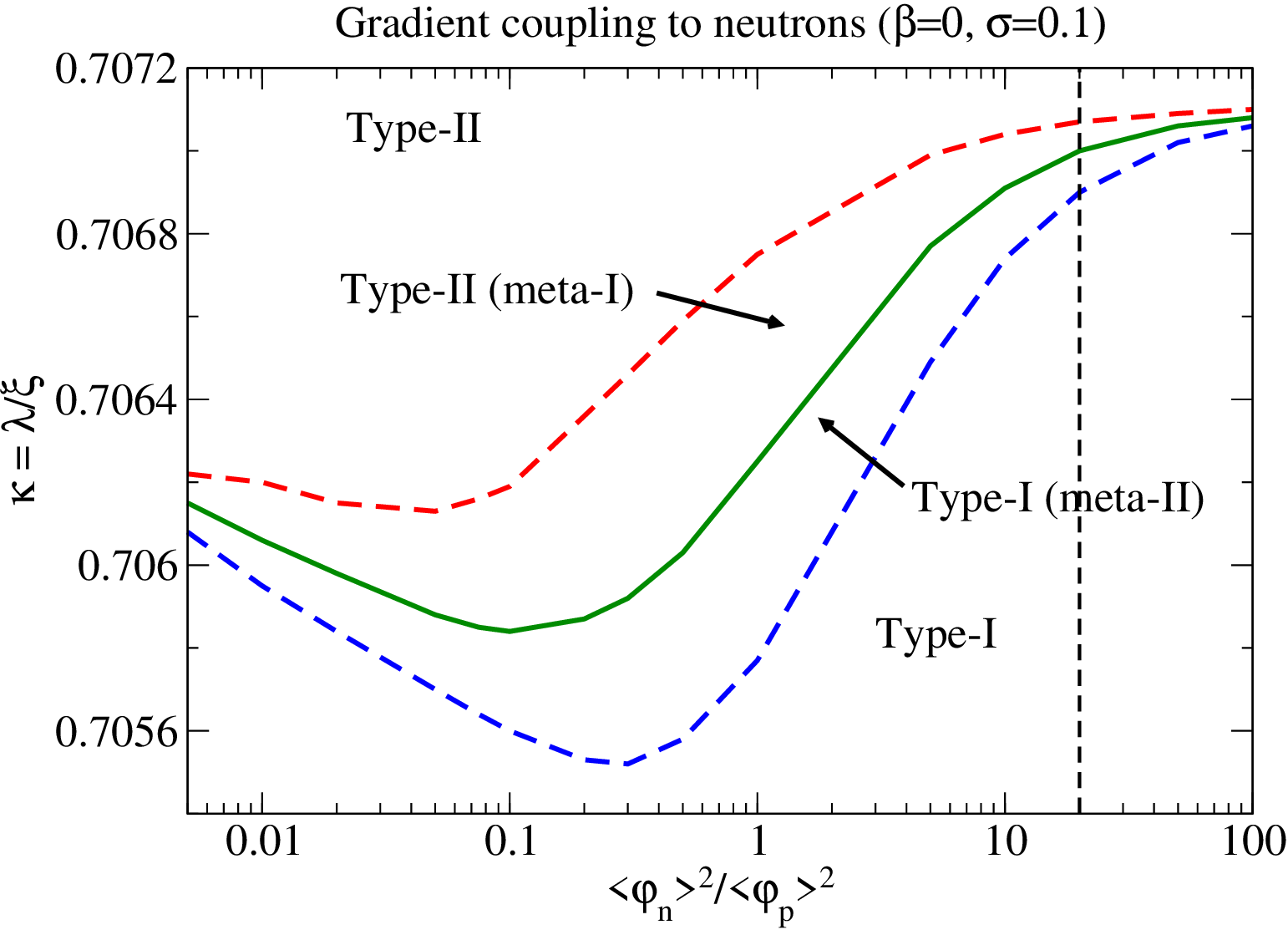}
\caption{(Color online)
Phase diagrams in the $\kappa$ vs.~$\<\phi_n\>^2/\<\phi_p\>^2$ plane.
Vertical dashed lines show $\<\phi_n\>^2/\<\phi_p\>^2=20$, the value used for 
other figures in this paper.
The left panel is for density coupling $\beta = -0.1$, but no
gradient coupling ($\sigma = 0$). The right panel is for
gradient coupling $\sigma = 0.1$, but no density coupling ($\beta = 0$).
In both cases, we see that the type-I/type-II transition converges to 
$\ka=1/\sqrt{2}$ as the neutron condensate disappears.
For the case of a density coupling, as the neutron condensate decreases,
the type-I/type-II boundary changes at $\<\phi_n\>^2/\<\phi_p\>^2\sim 10$
from a narrow region of type-II(n) phase
bands (thick line) to wider metastable regions.
}
\label{fig:phase-kappa-rho}
\end{figure}

\subsection{Phase diagrams}
\label{sec:phase-diagrams}

Figures \ref{fig:phase-kappa-beta}--\ref{fig:phase-kappa-rho} 
illustrate the additional structure in the phase diagram of the
superconductor induced by the couplings to a superfluid.
Each diagram is a two-dimensional slice through the
parameter space. 

Figure \ref{fig:phase-kappa-beta} shows the consequences of
a density coupling $\be$ between the
superfluid and superconductor. We see that the density coupling,
irrespective of its sign,
favors type-I superconductivity, pushing the
the critical $\ka$ for the type-I/type-II transition
up to higher values, forming a parabolic phase boundary
in the $\be$-$\ka$ plane, as expected from
\eqn{dk-beta}.
This can be thought of as
arising from the fact that nonzero $\be$
lowers the energy per flux
of the core of large flux tubes (see \eqn{beta-En}), 
which favors type-I superconductivity.

In the right panel we zoom in on the transition line
near $\be=0.5$ to show the substructure in the phase transition region
that is invisibly small in the left panel. As one would expect from
our discussion of Figure \ref{fig:energyperflux} (upper right panel),
on the type-II side of the transition there is a series of bands
distinguished by the number of flux quanta $n$ in the energetically favored
flux tube. ``Type-II ($n=1$)'' is the standard type-II superconductor.
With decreasing $\ka$ we find transitions to Type-II ($n=2$),
Type-II ($n=3$), 
and on up to $n=\infty$ which is a type-I superconductor.

In Figure \ref{fig:phase-kappa-sigma} we show the consequences of a
gradient coupling $\si$ between the superfluid and superconductor. We
see that the gradient coupling, irrespective of its sign, favors
type-II superconductivity, pushing the critical $\ka$ for the
type-I/type-II transition down to lower values, forming an inverted
parabolic phase boundary in the $\si$-$\ka$ plane, as expected from
\eqn{dk-sigma}.  It also makes the phase transition first order, with
spinodal lines where the unfavored phase becomes metastable. Both
these effects arise from the lowering of the energy of the wall of the
vortex, as explained in Sec.~\ref{sec:En-gradient}.

In Figure \ref{fig:phase-kappa-beta-sigma0.5} we show phase diagrams
for the combination of both density and gradient couplings,
fixing $\si=0.5$ and varying $\be$. As discussed in Sec.~\ref{sec:sign},
we expect that when $\si\neq 0$ the $\be\to -\be$ symmetry is
now broken. In the right panel we magnify
the transition region near $\beta = 0.5$, illustrating that on the
type-II side as $\ka$ decreases 
the number of flux quanta in the favored flux tube jumps from
1 to a finite value $n=5$, and then there is a sequence of bands in which
$n$ rises, reaching infinity when the superconductor becomes type I.
This is the expected behavior, based on 
our discussion in Sec.~\ref{sec:En-both}.

Finally, in Figure \ref{fig:phase-kappa-rho}, we anticipate one
direction in which this work could be extended, by
exploring the consequences of varying the ratio of the
superfluid density to the superconductor density, which up to now
was fixed to $\vevn^2/\vevp^2=20$, 
an appropriate value for neutral
beta-equilibrated nuclear matter, of the type we expect to find inside
neutron stars.
Figure \ref{fig:phase-kappa-rho} shows phase diagrams in the plane of
$\kappa$ and $\vevn^2/\vevp^2$ for a system with a density coupling
(left panel) and with a gradient coupling (right panel).

For the case of a density coupling 
we use a negative value of the coupling,
because this corresponds to an attractive
interaction, which gives smooth behavior in the limit where the
neutron condensate disappears, $\vevn^2/\vevp^2\to 0$.
As is clear from the plot, the type-I/type-II transition then converges
to the standard value for a single-component
superconductor, $\ka=1/\sqrt{2}$. 
For a repulsive interaction, the $\vevn^2/\vevp^2\to 0$
limit is singular: we discuss this in more detail below.
It is interesting to note that the effects of the density coupling
change dramatically with the relative densities of the neutrons and protons.
At $\vevn^2/\vevp^2\gtrsim 10$ the density coupling produces a thin region of
multi-flux-quantum ``type-II(n)'' phases, as was illustrated in
Fig.~\ref{fig:phase-kappa-beta}. But for lower values, it has a
similar effect to a gradient coupling, inducing metastable regions
on either side of the type-I/type-II boundary. 
This should be understandable in terms of the dependence of the coefficients
$b_\subhalf$ and $b_1$ (Eqn.~\eqn{eqn:dE-beta}) on $\vevn^2/\vevp^2$.
In Sec.~\ref{sec:En-density} we argued that if $b_1$ is large enough
then the $E_n/n$ curve has a minimum at finite $n$, yielding a
type-II(n) phase. We conjecture that as  $\vevn^2/\vevp^2$ gets smaller, 
$b_1$ becomes sufficiently negative that this is no longer the case,
and instead there is a maximum, leading to metastability of the
$n=0$ and $n=\infty$ states in spinodal regions around the
type-I/type-II boundary. This is a topic for future investigation.

For the case of a gradient coupling (right panel of 
Fig.~\ref{fig:phase-kappa-rho}) the effects of varying 
$\vevn^2/\vevp^2$ are less dramatic. It is interesting that, as for
a density coupling, the variation is non-monotonic. Again, we conjecture 
that this could be understood in terms of variation of the 
coefficients $s_\subhalf$ and $s_1$ (Eqn.~\eqn{eqn:dE-sigma}) with 
$\vevn^2/\vevp^2$. As the superfluid density drops to zero, its
effects become negligible, and the critical value of $\ka$
converges towards $1/\sqrt{2}$ as one would expect.

Finally, we discuss the singularity of the $\vevn^2/\vevp^2\to 0$
limit for a positive $\be$, i.e.~a repulsive 
density coupling between the neutron
and proton condensates.  From \eqn{V-rel-to-bulk}
we see that the expectation value of the neutron condensate is
$\vevn + \half \beta(\vevp - \phi_p)$, so far from the flux tube,
where $\phi_p$ is $\vevp$, it is $\vevn$. But in the core of the condensate
it is larger (there is less proton condensate to repel it). In fact, even
if the parameter $\vevn^2$ were zero or slightly negative, 
there would be a positive
neutron condensate in the core of the flux tube. This shows that for
positive $\be$ the neutrons do {\em not} decouple and become irrelevant
in the limit $\vevn\to 0$. We note two consequences of this.
Firstly, for small
$\vevn$ the $\be\to-\be$ symmetry discussed in Sec.~\ref{sec:sign}
is no longer present, because the effect of the flux tube on the neutron
condensate is no longer a small perturbation. Secondly,
in a system where $\vevn^2$ is small and negative (i.e.~the neutrons
just barely fail to condense in the presence of the proton condensate)
flux tubes could have superfluid cores, which is another topic that we leave
for future investigation.

\section{Conclusion}
\label{sec:conclusion}

We conclude that coupling a superconductor to a co-existing superfluid
causes significant modification of the energetics of the flux tubes.
On the basis of calculations restricted to the cylindrical geometry of
$n$-quantum flux tubes, we conclude that a coupling between the
densities of the condensates shifts the type-I/type-II boundary to
larger $\ka$, and, if the superfluid density is high enough,
appears to create an infinite number of new ``type-II(n)''
phases whose most stable flux tubes contain multiples of the basic
flux quantum.  A gradient coupling between the condensates leads to
metastable regions surrounding the transition between type-I and
type-II superconductivity.

As discussed in Section
\ref{sec:stability}, our calculation
corresponds to comparing the energy at zero and infinite separation
of flux tubes with varying numbers of flux quanta.
This leaves open the possibility that there might be additional
minima at finite separation. It is therefore possible that
in parts of the phase diagram there might be a different phase
from the ones we identify, namely an alternative type
of type-II superconductor in which the spacing between flux tubes
is fixed by the microscopic physics rather than by the strength of
the applied field. To resolve this question will require calculation
of the free energy of a pair of flux tubes at arbitrary separation.
Such calculations have been performed
for large separation \cite{Jacobs_Rebbi,Speight:1996px,Bettencourt:1994kf}, 
and
by perturbing about the Bogomolnyi point \cite{Mohammed}
and by numerical computation \cite{Hove:2002}.
In particular, the numerical methods that have been used 
recently to follow the interaction
and annihilation or vortex-antivortex pairs \cite{Gleiser:2007te}
would be readily applicable to the simpler time-independent
calculation of the interaction potential of flux tubes.
Another natural generalization of our calculation would be to allow
for non-$s$-wave pairing, such as the ${}^3P_2$ pairing that is
believed to occur in the neutron superfluid in the core of a neutron star.

Our results add another example to the class
of two-component Ginzburg-Landau models with non-standard
superconducting behavior.
Previous work in this area includes the $SO(5)$ model of high-temperature 
superconductivity, which has flux tubes described by a two-component
GL model, where each component
carries a different $U(1)$ charge, and only one of them condenses
in the vacuum \cite{MacKenzie:2003jp}.
Another example is the case of a two-component GL model where
both components have electric charge, very different mass, and nearly
the same Fermi energy. This system was found to have non-monotonic
$E(n)/n$ and intermediate minima in the interaction potential
\cite{Babaev:2004hk}. 

The exotic phenomena that we predict are localized to the
region around the type-I/type-II transition, so
they may not turn out to be relevant for the
inner core of a neutron star, which is believed to be
well inside the type-II regime \cite{BPP}.
However, given the extremely impressive recent progress 
in creating exotic systems such as multi-component
superfluids of trapped cold atoms, 
it seems quite conceivable that a material that is
both a superconductor and a superfluid might be created in the
laboratory, and could be studied under controlled conditions.
Our results would be directly relevant to such a material.

\section{Acknowledgements}
We thank Egor Babaev, Greg Comer, Igor Luk'yanchuk, 
Fidel Schaposnik, and Martin Speight for valuable discussions.
This research was
supported in part by the Offices of Nuclear Physics and High
Energy Physics of the U.S.~Department of Energy under contracts
\#DE-FG02-91ER40628,  
\#DE-FG02-05ER41375. 

\end{document}